\documentclass[fleqn,12pt,twoside]{article}
\usepackage[headings]{espcrc1}
\usepackage{amsmath}
\usepackage{amsfonts}
\usepackage{amssymb}
\readRCS
$Id: espcrc1.tex,v 1.2 2004/02/24 11:22:11 spepping Exp $
\usepackage{graphicx}

\newcommand{\ar}{\begin{eqnarray}}
\newcommand{\br}{\end{eqnarray}}

\newcommand{\BR}{{\mathbf{R}}}

\newcommand{\BRV}{{\langle} {\mathbf{R}}_F^2 (L){\rangle}}
\newcommand{\AVBR} { \overline{{\langle} {\mathbf{R}}_F^2 (L){\rangle}} }

\title{Localization of Polymers in Random Media: Analogy with
Quantum Particles in Disorder}

\author{Yadin Y. Goldschmidt \address{University of Pittsburgh, Department of Physics,
Pittsburgh, Pennsylvannia 15260.}
and Yohannes Shiferaw \address{University of California, Department of Medicine and
Cardiology, Los Angeles, CA 90095-1679.}}

\runtitle{Polymers in random media}
\runauthor{Y. Y. Goldschmidt and Y. Shiferaw}

\begin{document}

\maketitle

\begin{abstract}
\end{abstract}

\section{Introduction}

Polymers are very long chain-like macromolecules that play an
important role in a wide variety of physical systems.  Many of the
materials that we encounter in our every day lives, such as
plastics and rubber, are essentially a mesh like structure of
polymers. In biology, polymers also play a central role, as many
biological molecules, such as DNA, have a long chain-like
structure.  In this chapter we focus on the equilibrium
statistical mechanics of a single polymer chain immersed in a
quenched random medium.  For example, a chain that is free to move
in a porous material, such as a sponge or a complicated gel
network. This problem is relevant to a number of technologically
important processes, such as filtration \cite{rondelez,bishop},
gel permeation chromatography \cite{asher,slater}, and the
transport of polymers through porous membranes \cite{cannell}.
Here, we will not address these practical issues, but rather, give
an overview of the theoretical aspects of the problem.

In this chapter we present analytical and numerical results on the
conformational statistics of polymers in quenched random media. We
first consider the problem of a Gaussian polymer chain in a random
potential characterized by short ranged correlations. This
simplified mathematical model of a real polymer is tackled
analytically using the replica method, and also via a mapping to
the equivalent problem of a quantum particle in a random
potential. Here, we will focus on the statistical properties of a
long polymer chain, where the free-energy landscape is complex and
possesses many meta-stable states. Using the path integral mapping
between the partition function of a Gaussian polymer chain and the
imaginary time Schr\"odinger equation, we show that the glassy
phase can be understood in terms of the localized eigenfunctions
of the Schr\"odinger equation with a random potential.
Furthermore, we show that the glassy behavior can also be
described analytically by a one-step replica-symmetry-breaking
(RSB) solution.  We explore the connection between the replica
solution and the eigenfunctions of the Schr\"odinger equation, and
show that the one-step RSB solution can be interpreted in terms of
the dominance of localized tail states.

We proceed to investigate the more realistic case of a chain
immersed in a sea of hard obstacles that are randomly distributed
in space. Prior to our work it was often assumed in the literature
that that this problem is equivalent to the random potential problem.
Using Flory type free energy arguments we elucidate the
similarities and differences between this case and that of a
random potential with short ranged correlations. In particular, we
show that the dependence of the polymer size on chain length can
exhibit three possible scaling regimes depending on the system
size.

Finally, we introduce a more realistic model that includes a
self-avoiding interaction between monomers of the polymer chain.
We show that in the limit of a very long chain, and when the
self-avoiding interaction is weak, the equilibrium chain
conformation consists of many blobs with connecting segments.
These blobs are situated in regions of low average potential, in
the random potential case, or in a region of low density of
obstacles in the random obstacles case. We also show that as the
strength of the self-avoiding interaction is increased relative to the
strength of the random potential, the polymer
chain undergoes a localization-delocalization transition, where
the chain is no longer bound to a particular region of the medium
but can easily wander around under the influence of a small
perturbation.

\section{The statistics of a Gaussian chain in a random potential}
\label{sec1}

\subsection{Path integral formalism}

A polymer is a collection of molecules, called monomers, which
interact with each other to form a long flexible chain.  For
example, a typical polymer like polyethylene consists of a chain
of roughly $10^5$ $CH_2$ molecules.  The large number of monomers
allows for a statistical description of macroscopic properties
which are independent of details on the monomer scale. The
simplest mathematical model of a polymer chain is referred to as
the Gaussian chain \cite{doi}.  In this model the polymer is
described by a position $d$-dimensional vector $\BR(u)$, where $u$
is a continuous variable which satisfies $0<u<L$, and which runs
along the contour of a chain with length $L$.  The probability of
finding a conformation $\BR(u)$ is given by
 \ar P[\BR(u)] =
{\mathcal{N}} \ \exp\left[-\frac{d}{2 b^2} \int_0^L du \left (
\frac{d\BR(u)}{du} \right)^2 \right], \br 
where ${\mathcal{N}}$ is a normalization constant, and where $b$ is the average bond
length. In this description the self-avoiding interaction of the
chain with itself is not taken into acount, and in the absence of
an external potential the conformation of the chain resembles the
trajectory of a random walker with a mean step $b$. Thus, the
average end-to-end distance of the chain satisfies
\begin{eqnarray}
 \langle \left({\bf R}(L)-{\bf R}(0)\right)^2 \rangle = L\ b^2.
\end{eqnarray}

The conformational statistics of a polymer chain will change if
the chain is placed in a random environment.  We can model the
effect of the random environment by introducing an interaction
energy
\ar E_{int}[\BR(u)]=\int_0^L du \ V(\BR(u)), \br
where $V(\BR)$ is the potential energy of a monomer at the position
$\BR$ due to the environment. The potential function $V(\BR)$ will
depend on the type of random medium that is being studied, and
will be discussed in more detail later.  The conformational
probability is now
\begin{equation}
P[\BR(u)] = {\mathcal{N}} \ \exp\left[-\frac{d}{2 b^2} \int_0^L du
\left ( \frac{d\BR(u)}{du} \right)^2 -\beta \int_0^L du \
V(\BR(u)) \right],
\end{equation}
where we multiplied the free Gaussian conformational probability
with the  Boltzmann factor $\exp{\left (-\beta E_{int} [\BR(u)]
\right)}$, with $\beta=1/kT$. Given 
the conformational probability
we can we write the partition sum (Green's function) for the paths
of length $L$ that go from ${\mathbf{R}}$ to ${\mathbf{R}}'$ as
\begin{equation}
Z({\mathbf{R}},{\mathbf{R}}';L)=\int_{{\mathbf{R}}(0)
={\mathbf{R}}}^{{\mathbf{R}}(L)={\mathbf{R}}'} [d{\mathbf{R}}(u)]
{\exp}(-\beta H), \label{Z}
\end{equation}
where
\begin{equation}
H=\int^L_0 du \left[\frac{M}{2} \left(\frac{d{\mathbf{R}}(u)}{
du}\right)^2 + V\left({\mathbf{R}}(u)\right)\right], \label{Ham}
\end{equation}
and where $M=d/(\beta b^{2})$.  All the statistical properties of
the polymer chain can be derived from this partition sum.

In this chapter we will also consider the effects of the volume of
the random medium i.e. the system size.  In order to incorporate
this effect in the partition function we include a harmonic term
 \ar V\left({\mathbf{R}}(u)\right) \rightarrow
V\left({\mathbf{R}}(u)\right)+\frac{\mu}{2}{\mathbf{R}}^2(u), \br
where the coefficient $\mu$ will be a measure of the available
volume to the polymer (A larger $\mu$ implies a smaller system volume).
The confining well is also important to
ensure that the model is well defined, since it turns out that
certain equilibrium properties of the polymer diverge in the
infinite volume limit $(\mu \rightarrow 0)$.

A quenched random medium, such as a rough surface or a frozen gel
network, is a complex structure that can in principle be modelled
by a complicated potential function $V(\BR)$. However, we will not
be interested in the physical properties of a polymer chain
immersed in a specific environment, but rather in an ensemble of
similar environments. Hence, we will have to specify instead the
probability distribution of the random potential $V(\BR)$.  Here,
we will consider random potentials that are taken from a Gaussian
distribution defined by
\begin{equation}
\langle V({\mathbf{R}}) {\rangle}=0,\;        \; {\langle}
V({\mathbf{R}})V({\mathbf{R}}'){\rangle}=f\left(({\mathbf{R}}-{\mathbf{R}}')^2\right)
\label{VV} \ .
\end{equation}
In particular we will consider a correlation function of the form
\ar f\left( (\BR-\BR')^2) \right) = \frac{g}{(\pi \xi^2)^{d/2}}
\exp\left({- ({\mathbf{R-R'}})^2/\xi^2}\right),
\label{Gaussiancor1} \br
where $g$ determines the strength of the
disorder and the parameter $\xi$ conveniently controls the
correlation range of the random potential.  Here, we will consider
only the case of short-range correlations, where $\xi$ is much
smaller than the system size.

All the statistical properties of the polymer will depend on the
partition sum.  For instance, the average end-to-end distance of a
polymer chain that is free to move is given by
\begin{equation}
\overline{\langle  {\mathbf{R}}_F^2 (L) \rangle} =\overline{
\left(   \frac {\int d{\mathbf{R}}
d{\mathbf{R'}}({\mathbf{R}}-{\mathbf{R}'})^2
Z({\mathbf{R}},{\mathbf{R'}};L)}{\int d{\mathbf{R}} d{\mathbf{R'}}
Z({\mathbf{R}},{\mathbf{R'}};L)} \right)}, \label{rf2av}
\end{equation}
where the over-bar stands for the average of the ratio over
realizations of the random potential.  This average is referred to
as a quenched average, as opposed to an annealed average, where
the numerator and denominator are averaged independently.  In some
previous studies it has been argued that for the mean end-to-end
distance, as defined in Eq. \ref{rf2av}, one can replace the
quenched average by the more analytically tractable annealed
average.   However, this replacement can be justified only when
the system size is strictly infinite, since only in that limit can
the polymer sample all of space and find the most favorable
potential well that will be similar to its environment in the
annealed case.  The main problem with this approach is that in
practice we always deal with finite-size systems, and it is not
always easy to assess how big the system size has to be so that
the annealed average is a good approximation to the quenched
average.  In addition, the time it takes the chain to sample a
large volume is exceedingly long and unreachable over a reasonable
experimental time.

Before ending this section we would like to point out the
relationship between the statistical properties of a polymer chain
and a variety of other physical problems. The partition sum, as
written in Eq. (\ref{Z}), is in the form of a path integral over
all possible chain conformations. Since many physical problems can
be formulated in terms of path integrals, we can map the problem
of a Gaussian polymer in a random media to a wide variety of
seemingly unrelated problems.  First, we can map the partition sum
of a polymer chain to the density matrix of a quantum particle.
The mapping \cite{feynman,gold2} is given by
\begin{equation}
\beta \rightarrow 1/\hbar, \; \; L \rightarrow \beta \hbar.
\end{equation}
The density matrix of a quantum particle at inverse temperature
$\beta$ is then related to the partition function of the polymer
as
\begin{eqnarray}
 \rho(\BR,\BR';\beta)=Z(\BR,\BR';L=\beta \hbar,\beta=1/\hbar).
\end{eqnarray}
The monomer label $u$ is now interpreted as the Trotter
(imaginary) time, and $M$ as the mass of the quantum particle. The
density matrix is relevant to the equilibrium statistical
mechanics of a quantum particle, such as an electron in a dirty
metal.

The polymer partition sum for a Gaussian chain in random potential
can also be mapped into the partition sum of a flux-line in
type-II superconductors in the presence of columnar disorder
\cite{nelson,gold3}. Here, $\BR(u)$ is the transverse displacement
of the flux-line (or vortex). The variable $u$ is interpreted as
the distance along the $z$-axis (direction of the magnetic field),
and the variable $M$ corresponds to the line tension of the
flux-line. Thus a flux-line in three dimensions propagating along
the $z$-direction maps into a polymer in two spatial dimensions
(its projection onto the plane), and the columnar disorder maps
into point disorder. Although, in the case of several flux-lines
there is a repulsive electromagnetic interaction among them which
is not present for polymer chains.  In Table \ref{mapping} we have
summarized the relationships between a polymer chain, a quantum
particle, and a flux-line in a superconductor.

\begin{table}[ptb]
\begin{tabular}
[c]{|l|l|l|}\hline Polymer chain & Quantum particle & Flux line\\
& &  \\\hline
$u$  & $\tau$  & $z$   \\
(monomer label) & (imaginary time)&(plane label)\\\hline $\beta$ &
$1/\hbar$ & $\beta$\\\hline $L$ & $\hbar \beta$ & $L_z$\\
(chain length) & \ \  & (distance along z-direction)\\ \hline
$d/\beta b^2$  & $m$ & $\epsilon_\ell$\\
($b$=bond length) & (mass) & (line tension) \\\hline
\end{tabular}
\caption{The relationship between different physical systems.}
\label{mapping}
\end{table}

Finally, we can also map the polymer problem to the problem of diffusion
in a random catalytic environment \cite{nattermann,ebeling}.
This process describes, for instance,
auto-catalytic chemical reactions in a disordered background, or the
spreading of a population with a growth rate that depends on local random
conditions.  If we make the replacements
\ar
\frac{1}{2M\beta} \rightarrow D  \ ,
\ \beta V(\BR) \rightarrow -U(\BR) \ , \
L \rightarrow t
\br
where $D$ is the diffusion constant, and $U(\BR)$ is the growth rate
(or reaction rate) at position $\BR$, and $t$ is the time.  Then
$Z(\BR',\BR;t)$ gives the concentration of the constituents
at position $\BR'$ at time $t$, given a delta function concentration
concentrated at $\BR$ at time $t=0$.

\subsection{Flory arguments}

A remarkably fruitful approach to the statistics of polymers is
via simple and intuitive free energy arguments.  This approach was
used with great success by Cates and Ball \cite {cates} to derive
the essential scaling properties of a polymer in a random
potential.  Here, we  reproduce their beautiful intuitive
arguments in order to elucidate the effect of a finite volume on
the behavior of an untethered free chain in a random potential.
First, Cates and Ball argue that a Gaussian chain situated in an
infinite random medium is always collapsed in the long-chain
limit. Their argument goes as follows: Consider a white-noise
random potential $V(x)$ of zero mean whose probability
distribution is
\begin{equation}
P(V(x))\ \propto \ g^{-1/2}\ \exp (-V^2/2g).  \label{random}
\end{equation}
If we now coarse-grain the medium and denote by $\overline{V}$ the
average value of the potential over some region of volume
$\Omega$, then the coarse-grained potential will have the
distribution
\begin{equation}
P_{\Omega}(\overline{V})\ \propto \ (g/\Omega)^{-1/2}\ \exp
(-\Omega\overline{V}^{2}/2g). \label{coarse}
\end{equation}
Now, consider a polymer chain situated in the random potential,
and assume that it shrinks into a volume $\Omega$ corresponding to
a place where the mean potential $\overline{V}$ takes on a lower
value than usual. In this situation the free energy of the chain
is crudely estimated to be ( neglecting all numerical factors):
\begin{equation}
F(\Omega,\overline{V})=L/R^2+L\overline{V}+\Omega\overline{V}^2/2g.  \label{fe}
\end{equation}
Here, $L$ is the length of the chain (number of monomers), $R$ is
the radius of gyration (or end-to-end distance) and the volume
$\Omega$ is related to $R$ via $\Omega\sim R^d$ in $d$-spatial
dimensions. The first term on the r.h.s. is an estimate of the
free energy of a long chain confined to a region of size $R$ in
the absence of an external potential (see e.g. \cite{degennes},
Eq. I.12). The second term is just the potential energy of the
chain in the random potential of strength $\overline{V}$. The
third term arises from the chance of incurring a random potential
of strength $\overline{V}$. The quantity $\ln P(\overline{V})$
gives an associated effective entropy for the system. Minimizing
this free energy over both $\overline{V}$ and $\Omega$
determines the lowest free energy configuration. Minimizing with respect to $%
\overline{V}$
yields $\overline{V}=-Lg/\Omega$. Substituting in $F$ gives:
\begin{equation}
F(R)=\frac L{R^2}-\frac{L^2g}{2R^d}.  \label{fran}
\end{equation}
This shows that for any $d\geq 2$, $F\rightarrow -\infty $ as
$R\rightarrow 0 $. Thus, the mean size of the chain is zero, or in
the presence of a cutoff, the size of one monomer i.e.
\begin{equation}
R\ \sim 1,\ \ \ d\geq 2.  \label{Rg2}
\end{equation}
For $d<2$, the free energy has a minimum for
\begin{equation}
R\sim (Lg)^{1/(d-2)}\ , \ \ d<2\ ,  \label{Rl2}
\end{equation}
which in the long chain limit ($L\geq 1/g$) cuts off again at $R\ \sim 1$.
These results are the same as those for the case of an annealed potential
that is able to adjust locally to lower the free energy of the system. The
reason is that for an infinite system containing a finite (even though long)
chain, space can be divided into regions containing different realizations
of the potential, and the chain can sample all of these to find an
environment arbitrarily similar to that which would occur in the annealed
situation.

These results stand in contrast to the replica calculation of Edwards and
Muthukumar (EM) \cite{edwards}, who found that for a long chain
\begin{equation}
R\sim g^{-1/(4-d)},\ \ \ \ d<4  \label{REM}
\end{equation}
when $g^{2/(4-d)}L\rightarrow \infty $ , whereas $R\sim L^{1/2}$ when $%
g^{2/(4-d)}L\rightarrow 0$. Note that the result (\ref{REM}) is
independent of $L$ as opposed to Eq. (\ref{Rl2}). To reconcile the
two apparently different results, Cates and Ball argue that the
quenched case is different from the annealed case only for the
case when the medium has a \textit{ finite }volume $\cal V$. In a
finite box, arbitrarily deep potential minima are not present.
Instead the most negative $\overline{V}$ averaged over a region of
volume $\Omega \ll {\cal V}$ occupied by the chain, is
approximately (keeping only leading terms in the volume $V$) given
by solving the equation (the l.h.s. of which represents the area
under the tail of the distribution)
\begin{equation}
\int_{-\infty }^{\overline{V}}dy\ P_{\Omega}(y)\simeq \frac{\Omega}{{\cal V}}\ ,  \label{ieq}
\end{equation}
which yields
\begin{equation}
\overline{V}=-\sqrt{\frac{g\ln {\cal V}}{\Omega}}.  \label{vmin}
\end{equation}
This expression when plugged into Eq. (\ref{fe}) leads to (note
that the last term in (\ref{fe}) just becomes a constant
independent of $R$)
\begin{equation}
F(R)=\frac{L}{R^{2}}-L\sqrt{\frac{g\ln {\cal V}}{R^{d}}}.  \label{FQ}
\end{equation}
When this free energy is minimized with respect to $R$ it gives rise to
\begin{equation}
R\sim (g\ln {\cal V})^{-1/(4-d)} , \ \ \ \ d<4 \ \ \ . \label{RV}
\end{equation}
Using this value the binding energy per monomer becomes
\begin{eqnarray}
U_{bind}/L= (g \ln {\cal V})^{2/(4-d)}.
\label{bindingenergy}
\end{eqnarray}
For the polymer to be localized its total binding energy has to be
greater than the translational entropy $\ln {\cal V}$, which is
always satisfied for  $2<d<4$ when $L$ or ${\cal V}$ are large
(and for $d=2$ when $L$ is large).

\subsection{The relation to the localization of a quantum particle}
In order understand the conformational statistics of a Gaussian
chain in a random potential, we map the partition sum to an
imaginary time Schr\"odinger equation. This mapping (see
Ref.~{\cite{feynman}} Eqs.~(3.12)-(3.18)) is given by
\begin{equation}
Z({\mathbf{R}},{\mathbf{R}}';L)=\int_{{\mathbf{R}}(0)={\mathbf{R}}'}^{{\mathbf{R}}(L)={\mathbf{R}}}
[d{\mathbf{R}}(u)] {\exp}\left(-\beta H[{\mathbf{R}}(u)]\right)=
{\langle} {\mathbf{R}}|{\exp}(-\beta
L\hat{H})|{\mathbf{R}}'{\rangle}, \label{map}
\end{equation}
where
\begin{equation}
\hat{H}=-\frac{1}{2M\beta^2}\frac{\partial^2}{\partial
\hat{{\mathbf{R}}}^2}+\frac{\mu}{2}\hat{{\mathbf{R}}}^2
+V(\hat{{\mathbf{R}}}). \label{hamiltonian}
\end{equation}
So for a given realization of the random potential the polymer
partition sum can be expressed as a matrix element of the
imaginary-time evolution operator.  The matrix elements can be
expanded in eigenfunctions of the Hamiltonian operator to yield
\begin{equation}
{\langle} {\mathbf{R}}|{\exp}(-\beta
L\hat{H})|{\mathbf{R}}'{\rangle}= \sum_{m=0}^\infty {\exp}(-\beta
L E_m) \Phi_m({\mathbf{R}}) \Phi_m^*({\mathbf{R}}'), \label{eigen}
\end{equation}
where
\begin{equation}
\hat{H}\Phi_m({\mathbf{R}})=E_m \Phi_m({\mathbf{R}}).
\label{eigenvalue}
\end{equation}

The Schr\"odinger equation with a random potential is a well known
problem that has been intensely studied for a long time
\cite{anderson,review1,frohlich,lifshits}. The main property that
we will use is that when $V({\mathbf{R}})$ has short range
correlations (i.e. correlation length is shorter than any other
length scale in the problem), and if the system size is infinite,
then in any dimension all eigenstates with energy below a critical
energy $E_M$ ( referred to as the mobility edge) are exponentially
localized in the form
\begin{equation}
\Phi_m ({\mathbf{R}}) \sim \exp{(-
|{\mathbf{R}}-{\mathbf{R}}_m|/\ell_m)}. \label{localized}
\end{equation}
Here ${\mathbf{R}}_m$ is the localization center of the $m^{th}$
state, and $\ell_m$ is the localization length of that state. The
localization length satisfies $1/\ell_m = \beta\sqrt{2M|E_m|}$ for
$E_m \ll 0$, i.e. deep in the tail region.  For $E>E_M$ extended
states exist when $d>2$. For $d=1,2$ there is no mobility edge and
all states are exponentially localized. The states with energies
$E>E_M$ are called extended since they are no longer localized but
are spread over a finite fraction of the system. Also, it is known
that the eigenvalues of the localized states are discrete, while
the eigenvalues of the extended states form a continuum.

For finite system size, or if $\mu \neq 0$ in the Hamiltonian
given in Eq.~(\ref{hamiltonian}), the above discussion has to be
modified.  First, the eigenfunctions are always discrete in any
dimensions. But even in one dimension as the energy increases the
width of the localized states eventually becomes comparable to the
system size and thus a localized particle of that energy can go
from one end of the sample to the other. Thus the distinction
between localized and extended states becomes blurred for a finite
system at energies much above the ground state. Nevertheless,
there will still be a  qualitative difference between the low
energy tail states and the higher energy states with large
localization lengths.

All the physical properties of the polymer chain can be expressed
in terms of the eigenstates of Schr\"odinger equation. For
instance we can write the end-to-end distance for a given
realization of the random potential as
\begin{equation}
\BRV_V=\frac{ 2\sum_{m} \left( a_m \int d\BR \BR^2 \Phi_m^*(\BR)
-\left| \int d\BR \BR \Phi_m(\BR) \right|^2 \right) \exp(-\beta L
E_m) }{\sum_{m} |a_m|^2 \exp(-\beta L E_m)}, \label{exprf}
\end{equation}
where $a_m = \int d\BR \Phi_m(\BR) $, and where $\langle \cdot
\rangle_V$ refers to a configurational
average for the case of a
fixed realization of random potential. When $L$ is large enough so
that $(E_{1}-E_{gs})L>>1$, where $E_1$ is the eigenvalue of the
first excited state and $E_{gs}$ is the ground state eigenvalue,
then only the ground state contributes. In this case we have
\begin{equation}
\BRV_V=2\frac{\int d\BR \BR^2 \Phi_{gs} (\BR)}{\int d\BR
\Phi_{gs}(\BR)}
 -2\left( \frac{\int d\BR \BR \Phi_{gs}(\BR)}{\int d\BR \Phi_{gs}(\BR)}
\right)^2 ,
\end{equation}
where $\Phi_{gs}(\BR)$ is the ground state eigenfunction.  It can
be shown that the ground state wave function is positive definite
and so in the large $L$ limit $\BRV^{1/2}$ can be interpreted as
the width of the ground state eigenfunction.  Assuming the ground
state has the form given in Eq.~(\ref{localized}), we can write
$\BRV_V=2d(d+1)\ell_{gs}^2$, where $\ell_{gs}=\ell_0$ is the
localization length of the ground state. Upon averaging over all
realizations of the random potential we get that
$\AVBR=2d(d+1)\overline{\ell_{gs}^2}$, and so the quenched average
of the end-to-end distance, in the long chain limit, is
proportional to the square of the average localization length of
the ground state eigenfunction.

\subsection{Localized eigenstates and glassy behavior}

\begin{figure}
\begin{center}
\includegraphics[width=6cm]{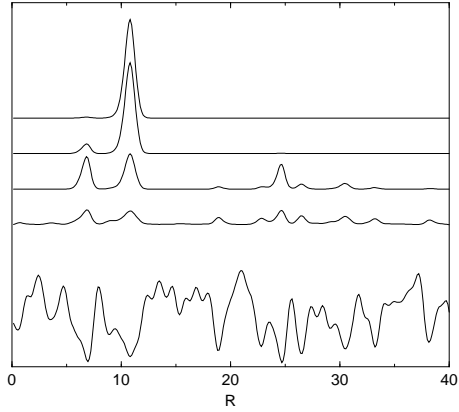}
\end{center}
\caption{Plot of $P(R,L)$ vs $R$ for four values of $L$. The
bottom most curve is the random potential sample that is used.
From bottom to top we use $L=.3,1,10,20$. The model parameters are
$M=1/2$, $g=25$, $\beta=1$, and $\mu=0.01$. We use a lattice of
size S=40 with $\Delta=0.2$.  The random potential is modeled by
generating $N$ numbers $\{V_{\xi}(i)\}_{i=1,...,N}$ which satisfy
$\langle V_{\xi}(i) V_{\xi}(i+l) \rangle \propto \exp (-\Delta^2
l^2/\xi^2)$, where $\xi=1/\sqrt{2}$. } \label{dist}
\end{figure}

Using the path integral mapping we can evaluate the partition sum
by solving the discretized Schr\"odinger equation.  In $d=1$ this
can be accomplished by simply diagonalizing an $N\times N$
Hamiltonian matrix, with lattice spacing $\Delta=S/N$, where $S$
is the system size.  Details of the numerical procedure are given
in \cite{shifgold}.   Using the lattice computation, we explore
the connection between the eigenstates of the Schr\"odinger
equation and the physical properties of the polymer chain.  Here,
we focus on the probability distribution defined as
\ar
P(\BR,L)=Z(\BR,\BR,L)/\int Z(\BR,\BR,L) d\BR
\br
which can be
interpreted as the probability of finding a closed polymer chain
of length $L$ which passes through the point $\BR$ (for a given
realization of the random potential). We consider this probability
distribution since it gives the most direct connection between the
chain properties and the eigenfunctions of the Schr\"odinger
equation.  In Fig.~\ref{dist} we plot $P(R,L)$ vs. $R$ for four
different chain lengths. We also include a  plot of the random
potential sample that is used. From the plot we can see clearly
that as the chain length is increased the probability distribution
tends to localize around a few valleys of the random potential
landscape. As $L$ is increased further there is only one peak as
the chain finds the most favorable position.  These results can be
explained in terms of the eigenfunctions of the Schr\"odinger
equation using the expansion
\begin{equation}
Z({\mathbf{R}},{\mathbf{R}};L)=\sum_{m} {\exp}(-\beta L E_m)
|\Phi_m({\mathbf{R}})|^2 . \label{loopexp}
\end{equation}
Which shows that as $L$ is increased the localized tail states
dominate the partition sum until only the ground state remains
when $(E_{1}-E_{gs})L>>1$.

\begin{figure}
\begin{center}
\includegraphics[width=6cm]{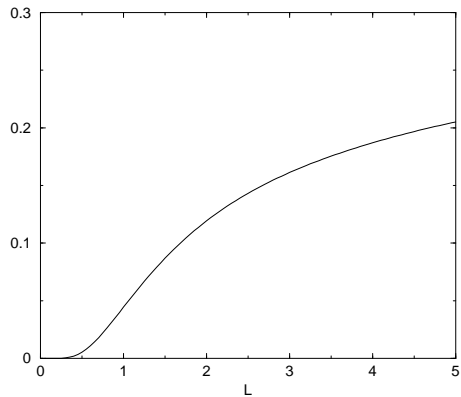}
\caption{Plot of $\Delta_F/\AVBR$ vs. $L$. The parameters are the
same as in Fig.~\ref{dist} and we average over $1000$ random
potential realizations.}
\end{center}
\label{delf}
\end{figure}

The dominance of tail states for long polymer chains leads to
large sample-to-sample variations in measured physical quantities.
This is because the partition sum, for a given realization of the
random potential, is dominated by a few favorable conformations
which are strongly sample dependent. In Fig. \ref{delf} we plot
the relative sample-to-sample fluctuations of the end-to-end
distance $\BRV$, as a function of chain length.  The relative
sample-to-sample fluctuation is defined as $\Delta_F/\AVBR$ with
\ar
\Delta_F=\left(\overline{\BRV^2}-\overline{\BRV}^2\right)^{1/2},
\br
and where the averaging is done over many realizations of the
random potential.  From the plot it is clear that as the length of
the chain is increased the sample-to-sample fluctuations increases
dramatically at a chain length $L_c \sim 0.5$.

Large sample-to-sample fluctuations in measured physical
quantities are typical of glassy systems, which are a broad class
of systems characterized by rugged energy landscapes \cite{MPV}.
This is precisely the case here, where the free energy of a long
chain will be dominated by conformations that reside in the deep
valleys of the random potential landscape.  The main new result
here is that the emergence of glassy characteristics of a polymer
chain in random media can be traced to the dominance of localized
tail states of the Schr\"odinger equation.

\subsection{Averaging over disorder: The replica trick}
\label{replica}

The numerical results in the previous section
reveal that a polymer in a random potential can be viewed as a
glassy system. In order to develop this point of view, we apply
the replica method, which has been successfully applied in the
study of glassy systems \cite{MPV}.  The replica approach was
invented in order to compute quenched averages such as $\AVBR$, as
given in Eq. (\ref{rf2av}), which is difficult since it entails
averaging a ratio of the partition sum.  The starting point is the
formal identity
 \begin{eqnarray}
\ln(Z)= \lim_{n \rightarrow 0} \frac{Z^n - 1}{n},
\end{eqnarray}
which can be used to calculate the quenched free-energy
\begin{eqnarray}
\overline{F}=-kT\ \overline{\ln (Z)} =-kT \lim_{n \rightarrow 0}
\frac{\overline{Z^n} - 1}{n},
\end{eqnarray}
where the order of the averaging and taking the $n \rightarrow 0$
limits has been interchanged, hopefully with no ill ramifications.
The average of a term like $Z^n$, where $n$ is a positive integer,
is easy to implement by creating $n$-identical copies of the
system, referred to as replicas, and then averaging over the random
potential. This process of averaging results
in interactions between different replicas. The resulting
analytical expression is then continued analytically to $n=0$.

Introducing $n$-copies of the system and computing
$\overline{Z^n}$ it is straight forward to show that the average
end-to-end distance can be formally written as
\ar
\AVBR=
 \lim_{n \rightarrow 0} \frac{
\int \prod_{a=1}^{n} d\BR_a  \prod_{a=1}^{n} d\BR'_a
 (\BR_1 -\BR'_1)^2
Z_n(\{{\mathbf{R}}_a\},\{{\mathbf{R}}'_a\};L) } { \int
\prod_{a=1}^{n} d\BR_a  \prod_{a=1}^{n} d\BR'_a
Z_n(\{{\mathbf{R}}_a\},\{{\mathbf{R}}'_a\};L) }, \label{wanderF}
\br
where
\ar
Z_n(\{{\mathbf{R}}_a\},\{{\mathbf{R}}'_a\};L)
=\int_{{\mathbf{R}}_a(0)={\mathbf{R}}_a}^{{\mathbf{R}}_a(L)={\mathbf{R}}'_a}
\prod_{a=1}^n[d{\mathbf{R}}_a]\exp(-\beta H_n), \label{Zn1}
\br
and where
\begin{eqnarray}
H_n=\frac{1}{2}\int_0^L du \sum_{a} \left[ M\left(\frac{d
{\mathbf{R}}_a (u)}{
du}\right)^2 +\mu {\mathbf{R}}^2_a (u) \right] \nonumber \\
-\frac{\beta}{2} \int^L_0 du \int^L_0 du' \sum_{ab} f\left(
({\mathbf{R}}_a (u)-{\mathbf{R}}_b (u'))^2 \right). \label{Hn1}
\end{eqnarray}
Thus, the averaged equilibrium properties of the polymer can be
extracted from an $n$-body problem by taking the $n\rightarrow 0$
limit at the end. This limit has to be taken with care, by solving
the problem analytically for general $n$, before taking the limit
of $n\rightarrow0$.

\subsection{The Replica Variational Approach}

In order to compute quenched averages of the polymer chain we will
have to solve the $n-$body replicated partition sum given in Eq.
(\ref{Hn1}).  This path integral cannot be evaluated analytically
and a variational approach has been used in Refs.
\cite{gold,gold2} to make further progress. The procedure is to
follow the work of Feynman \cite{feynman} and others
\cite{shakhnovich,MP} and model $H_n$ by a solvable trial
Hamiltonian $h_n$ which is determined by the stationarity of the
variational free energy
\begin{equation}
n \langle F \rangle = {\langle H_n - h_n \rangle}_{h_n} -
\frac{1}{\beta} \ln \int [d\BR_1] \cdots [d\BR_n] \exp(-\beta
h_n). \label{varfree1}
\end{equation}
Following \cite{gold} we use a quadratic trial Hamiltonian of the
form
\begin{eqnarray}
h_n=\frac{1}{2}\int_0^L du \sum_{a} \left[ M\left(\frac{d
{\mathbf{R}}_a (u)}{d
u}\right)^2 +\mu {\mathbf{R}}^2_a (u) \right] \nonumber \\
-\frac{1}{4L}\int^L_0 du \int^L_0 du' \sum_{ab} p_{ab} \left(
\BR_a (u) - \BR_b (u') \right)^2, \label{trialh}
\end{eqnarray}
where the matrix elements $p_{ab}$ are the variational parameters.
The physical motivation for this ansatz is that the
replica-replica interaction in the original Hamiltonian is modeled
by a quadratic interaction which can be different for different
replica pairs.  Also, the form of the quadratic interaction was
chosen specifically \cite{gold} to preserve the translational
invariance of the original Hamiltonian.
Apart from the $\mu$-dependent term, (or alternatively in the limit $\mu\rightarrow 0$),
the original Hamiltonian $H_n$ given in Eq. (\ref{Hn1}) is invariant
under the transformation
\begin{eqnarray}
{\mathbf  R}_{a}(u)\rightarrow {\mathbf  R}_{a}(u)+ {\mathbf C},\ \ \ a=1,\cdots,n
\label{sym}
\end{eqnarray}
where ${\mathbf C}$ is a constant vector. This reflects the fact
that in the infinite system, without a confining harmonic term,
after averaging over the random potential the interaction is
translational invariant. A good variational Hamiltonian must be
one that can preserve this translational symmetry. Thus the
variational ansatz must be rich enough to implement the relation
(\ref{sym}). The ansatz chosen by EM \cite{edwards} violated this
translational invariance for any $q \neq 0$, where $q$ denoted
their single variational parameter. This imposed an unphysical
origin on the system when none existed in the infinite volume
limit that they considered. It was actually shown in
Ref.~\cite{gold} that if the variational ansatz is rich enough the
translation invariance actually emerges from the variational
extremization even if not assumed explicitly for the trial
Hamiltonian.

\begin{figure}
\centering
\includegraphics[width=6cm]{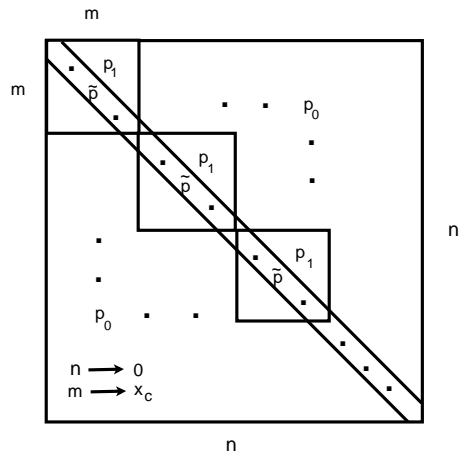}
\caption{Illustration of first step replica symmetry breaking.
The limits $n\rightarrow 0$ and $m\rightarrow x_c$ have to be taken
with $0<x_c<1$. }
 \label{1rsb}
\end{figure}

We will now assume that in the $n \rightarrow 0$ limit the matrix
$p_{ab}$ can be parameterized according to the 1-step replica
symmetry breaking scheme of Parisi \cite{MP}. In the 1-step RSB
scheme, the matrix $p_{ab}$ can be parameterized as
$(\tilde{p},p(x))$ with
\begin{eqnarray}
p(x)= \left\{ \begin{array}{l}
p_0 \ \ \ \ \ 0<x<x_c \  \\
p_1 \ \ \ \ \ x_c<x<1 \ ,
\end{array} \right.
\label{pmat}
\end{eqnarray}
and where $x$ is Parisi's replica index.  See Figure(\ref{1rsb})
for an illustration.

Calculating the variational free energy using Eq.~(\ref{varfree1})
it could be expressed \cite{gold} in the limit $n\rightarrow 0$ as a function of the
four variational parameters. i.e. $F=F(\tilde{p},p_0,p_1,x_c)$. The
stationarity of the free energy yields four coupled non-linear equations for
the four variational parameters. In the limit of small $\mu$ (large
volume) and long chains ($L$ large) these equations were solved analytically.
Denoting by $\lambda$ the combination
\begin{eqnarray}
\lambda=\mu-\tilde{p}+(1-x_c) p_1+x_c p_0,
\label{lambda}
\end{eqnarray}
we find from the analytical variational solution
\begin{eqnarray}
\lambda\sim\left(  g\ \left|  \ln\mu\right|  \right)  ^{4/(4-d)}%
.\label{lamfin}
\end{eqnarray}
From this final result we can obtain the radius of gyration which can
be shown to be proportional to $\lambda^{-1/4}$. Recalling also that
$|\ln \mu| \propto \ln{\cal V}$, we find
\begin{eqnarray}
R_g  & \sim\lambda^{-1/4}\sim\left(g\ \left|  \ln\mu\right|  \right)  ^{-1/(4-d)}
\sim\left(g\ \ln {\cal V}\right)
^{-1/(4-d)},
\label{rfin}%
\end{eqnarray}
a result which coincides with the prediction of Cates and Ball for the
case of a finite volume. It could also be shown that a
replica-symmetric ansatz for the matrix $p$ yields a result consistent
with the annealed average of the disorder where the chain collapses in
the limit of $L \rightarrow \infty$.

For non-asymptotic values of $\mu$ and $L$ the non-linear stationarity
equations equations could be solved
numerically \cite{shifgold} using a standard iterative method \cite{press}.
We found that for a given set of
parameters there is a chain length $L_c$ (which depends on the
strength of the disorder) such that for $0<L<L_c$ there is only a
replica symmetric solution.  This is the case when the variational
parameters satisfy $x_c=1$ and $s_0=s_1$. For $L>L_c$ there is
still a replica symmetric solution but we also find an additional
replica symmetry breaking solution. So in this regime we find an
additional solution such that $0<x_c< 1$ and $s_0 \neq s_1$.  In
order to decide which solution correctly describes the physics in
that regime  we compare their respective predictions to the
lattice computation of $\AVBR$.

\begin{figure}
\begin{center}
\includegraphics[width=6cm]{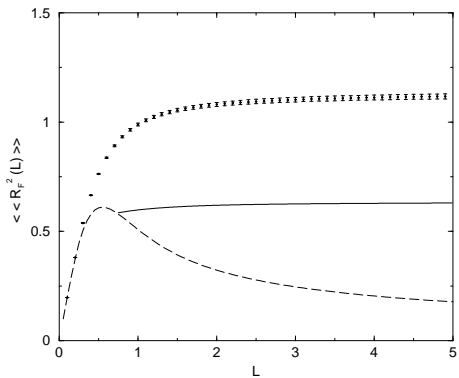}
\caption{Plot of $\AVBR$ vs. $L$. The dotted line is generated by
averaging over $10000$ samples, and error bars are found by
computing the standard deviation of $10$ sets of $1000$ samples.
The dashed line is the RS solution, and the solid line is the RSB
solution. } \label{rf1}
\end{center}
\end{figure}

In Fig.~\ref{rf1} we plot the mean squared displacement $\AVBR$
vs. $L$ for a given set of parameters. We plot this quantity using
the lattice result, and also using the two predictions of the
variational method. Note that in the labels of the plots the
average over the disorder is denoted by a second set of brackets
rather than an over-bar.
For $L$ below $L_c \approx 0.73$ there is only a RS solution which
is very close to the lattice prediction.  For $L$ greater than
$L_c$ the RS and RSB solutions are different and it is clear that
the RSB solution is closer to the lattice result. We can see that
the end-to-end distance saturates at a constant value as $L$
increases.  This behavior is correctly predicted by the RSB
solution but not by the RS solution.

\begin{figure}
\begin{center}
\includegraphics[width=6cm]{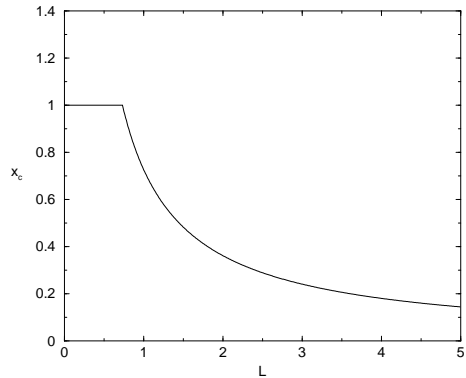}
\caption{Plot of $x_c$ vs. $L$.
The parameters are the same as those used in Fig.~\ref{rf1}.
\label{zplot} }
\end{center}
\end{figure}

So far we have seen that the glassy characteristics of a polymer
in a short range correlated random potential is closely related to
the dominance of low energy eigenfunctions.  We also know that the
variational solution possesses a RS solution for $L<L_c$ and an
RSB solution for $L>L_c$. It is well known that replica symmetry
breaking is typically associated with glassy behavior, and for our
model we show that the onset of RSB is precisely when the system
begins to exhibit glassy behavior.  The variational parameter that
best reveals the transition between RS and RSB is the break point
$x_c$.  If $x_c=1$ then there is only an RS solution, and if
$0<x_c<1$ then that corresponds to an RSB solution.  In
Fig.~\ref{zplot} we plot $x_c$ vs. $L$ using the same parameters
that were used in Fig.~\ref{rf1}.
We can see that onset of the RSB solution is at $L_c \approx .73$,
after which we find that the break point decreases like $x_c
\propto 1/L$. If we compare this result to the plot of
$\Delta_F/\AVBR$ vs $L$ in Fig.~\ref{delf}, we see that near $L_c
\approx 0.5$ the sample-to-sample fluctuations begin to rise
rapidly.  This result provides strong evidence that when the RSB
solution is valid the polymer chain does indeed exhibit glassy
behavior.

\subsection{Physical interpretation of the 1-step RSB solution}

 In this section we study the physical
interpretation of the replica symmetry breaking solution.  Our
purpose is to see if the underlying physical picture predicted by
1-step RSB  is indeed consistent with the presence of
exponentially localized eigenstates. The following analysis is
valid for a very long polymer (large $L$) when the system becomes
glassy. We begin by evaluating the replicated partition sum
defined as
\begin{equation}
\tilde{Z}_n(\{\BR_a\})=\int_{\BR_a(0)=\BR_a}^{\BR_a(L)=\BR_a}
\prod_{a=1}^n[d\BR_a]\exp(-\beta h_n), \label{Zrep}
\end{equation}
where $h_n$ is the quadratic trial Hamiltonian in
Eq.~(\ref{trialh}). Since $h_n$ is quadratic the path integrals
can be evaluated analytically and the final result can we written
in the form
\begin{equation}
\tilde{Z}_n(\{\BR_a\})={\rm const.} \times \exp \left(-\frac{1}{2}
\sum_{ab} Q_{ab}^{-1} \BR_a \cdot \BR_b \right). \label{ZQ}
\end{equation}
The details of this calculation along with the relationship
between the matrices $Q_{ab}$ and $p_{ab}$ are given
\cite{shifgold}. Now, since $p_{ab}$ was parameterized according
to the 1-step RSB scheme, it implies that $Q_{ab}$ can also be
parameterized in the same way.

Mezard and Parisi \cite{MP} discuss the interpretation of a
representation of the form (\ref{ZQ}) for the case of directed
polymers. In particular they show how to deduce the structure of
the probability distribution
\begin{eqnarray}
P_V(\BR)=\tilde{Z}_V (\BR,L)/\int d \BR \tilde{Z}_V(\BR,L),
\label{PV}
\end{eqnarray}
which is the probability of finding a polymer loop that passes
through $\BR$ for a given realization (which we denote by $V$) of
the the random potential. Here $\tilde{Z}_V (\BR,L)$ is just the
partition sum $Z(\BR,\BR;L)$ as given in Eq.~(\ref{Z}). This
probability is related to the replicated partition function given
in Eq.~(\ref{Zrep}) by
\begin{eqnarray}
P_V(\BR)= \lim_{ n \rightarrow 0} \int d\BR_2 \cdots d\BR_n\
\tilde{Z}_n(\{\BR_a\})_{\BR_1=\BR}. \label{PVrel}
\end{eqnarray}
Mezard and Parisi's analysis has to be adapted for the case of
real (non-directed) polymers of length $L$ in a random potential
which is independent of time. The changes will be pointed out
below.

If  $Q_{ab}$ is parameterized by $\{\tilde{q},q(x)\}$ such that
\begin{eqnarray}
q(x)= \left\{ \begin{array}{l}
q_0 \ \ \ x < x_c \  \\
q_1 \ \ \ x > x_c \ ,
\end{array}
\right.
\end{eqnarray}
one proceeds to obtain $P_V(\BR)$ by the following procedure:

\begin{flushleft} 1. For each sample ( a realization of the random potential)
generate a random variable $\BR_0$ which is picked from the
distribution
\begin{equation}
{\cal P} (\BR_0)=\frac{1}{(2\pi q_0)^{d/2}} \exp  \left(
-\frac{\BR_0^2}{2q_0}\right).
\end{equation}

2. Consider a set of ``states'' labeled by the index $\alpha$
whose physical meaning will be elucidated shortly. Each of these
states is characterized by a weight $W_\alpha$ and a position
variable $\BR_\alpha$. Given $\BR_0$, the variables $\BR_\alpha$
are an infinite set of uncorrelated random variables distributed
according to
\begin{equation}
{\cal P}(\BR_1,\BR_2,\cdots)=\prod_\alpha\frac{1}{(2\pi(q_1-q_0))^{d/2}}
\exp  \left( -\frac{(\BR_{\alpha}-\BR_0)^2}{2(q_1-q_0)}
\right). \label{positiondist}
\end{equation}
The distribution of weights will be discussed below.

3. Given these ``states'' for a given sample, The probability
distribution $P_V(\BR)$ for that sample has the form
\begin{equation}
P_V(\BR)=\sum_{\alpha} W_{\alpha}
\frac{1}{(2\pi(\tilde{q}-q_1))^{d/2}}
 \exp \left(
-\frac{(\BR-\BR_{\alpha})^2}{2(\tilde{q}-q_1)} \right).
\label{PVR}
\end{equation}
\end{flushleft}

The weights $W_\alpha$ are given in terms of some ``free energy''
variables $f_\alpha$:
\begin{eqnarray}
W_\alpha = \frac{\exp(-\beta f_\alpha)}{\sum_{\gamma} \exp(-\beta
f_{\gamma}) }. \label{falpha}
\end{eqnarray}
These free energy variables are chosen from an exponential
distribution
\begin{eqnarray}
P[f_\alpha] \propto \exp (x_c \beta f_\alpha) \ \theta(f-\bar{f}),
\label{distf}
\end{eqnarray}
where $\bar{f}$ is an upper cutoff.

What is the meaning of these variables in the present case? To
determine the weights $W_{\alpha}$ we compare Eq.~(\ref{PVR}) to
the eigenfunction expansion given in Eq.~(\ref{loopexp}). From
Eq.~(\ref{loopexp}) together with Eq.~(\ref{PV}) it becomes clear
that
\begin{equation}
P_V(\BR)=\sum_\alpha A_\alpha |\Phi_\alpha(\BR)|^2,
\end{equation}
where
\begin{equation}
A_\alpha = \frac{\exp(-\beta L E_\alpha)}{\sum_{\gamma}
\exp(-\beta L E_{\gamma}) }. \label{eigenweights}
\end{equation}
Comparing Eq.~(\ref{PVR}) and Eq.~(\ref{eigenweights}) it becomes
obvious that $W_\alpha=A_\alpha$ and
\begin{equation}
\Phi_\alpha^2(\BR) \propto  \exp \left(
-\frac{(\BR-\BR_{\alpha})^2}{2(\tilde{q}-q_1)} \right).
\end{equation}
Hence, the ``states'' labeled by $\alpha$ are in our case the
actual eigenstates of the imaginary time Schr\"odinger equation.
These are localized tail states centered at position
$\BR_{\alpha}$ with an associated ``weight'' $W_\alpha$. Thus the
1-step RSB solution approximates the tail states by a fixed
Gaussian form.

The width of these Gaussians, denoted by $w_{0}$, give an estimate
of the size of the polymer chain i.e. $w_{0} \sim R_F$.  In the
limit of large $L$ and small $\mu$ it can be shown to leading
order \cite{gold} that $w_0^2 \sim d/(2\beta \sqrt{\lambda M)}$,
where
\begin{equation}
\lambda=\frac{d^{4/(4-d)}}{(2\pi)^{2d/(4-d)}}\left(
\beta^{2}M\right) ^{(4+d)/(4-d)}\left(  g\ \left|  \ln\mu\right|
\right)  ^{4/(4-d)} \ .
 \label{lamdafin}
\end{equation}
So in terms of the disorder strength and the system size the chain
size scales like
\ar
R_F \propto (g |\ln \mu|)^{-1/(4-d)}\propto (g \ln {\cal V})^{-1/(4-d)} .
\label{scaling}
\br
It should be emphasized that the subtle dependence on the volume of
the system is a direct consequence of replica symmetry breaking.
In fact, as shown in Fig. \ref{rf1} the replica symmetric
solution does not correctly describe the size of the polymer
chain, since it fails to capture the dominance of localized tail
states.

We now consider the distribution $P(\BR_\alpha)$ given in
Eq.~(\ref{positiondist}). This is just the distribution for the
localization centers $\BR_\alpha$ for a given value of $\BR_0$.
Hence, we can calculate the average distance between the localized
states for a given sample.  We find that the width $w$ of the
Gaussian $P(\BR_\alpha)$ satisfies $w^2=d(q_1-q_0)$. For small
$\mu$ and large $L$ it is straight forward to show that that $w^2
\approx d/(\beta \mu Lx_c)$,
where the break point $x_c$ is given
by
\begin{eqnarray}
x_{c}=\frac{1}{L}\left(  \frac{d^{d-2}}{(2\pi)^{d}}g^{2}\beta^{d+4}%
M^{d}\left|  \ln\mu\right|  ^{d-2}\right)  ^{-1/(4-d)}.
\label{xcf}%
\end{eqnarray}

\subsection{Density of states and the 1-step RSB solution}

To develop the analogy further we first notice that the free
energies $f_\alpha$ are equal to $ L E_\alpha$. This make sense if
we think of $|E_\alpha|$ as representing the binding energy per
monomer, and thus $f_\alpha= L E_\alpha$ represent the total
energy of the chain.   These arguments lead us to expect that
within the 1-step RSB scheme, the energy variables $E_\alpha$ are
independent random variables taken from an exponential
distribution:
\begin{equation}
P[E_\alpha] \propto e^{\beta L x_c E_\alpha}
\theta(\overline{E}-E_\alpha) \ \ , \label{eigendist2}
\end{equation}
with $\overline{E}$ being some energy scale determined by the
upper cutoff of the tail region. We will now argue that the
distribution given above is just the expected distribution of
ground-state energies i.e. the probability of finding the lowest
energy level to have energy $E$. We first review some very basic
results of extreme value statistics as presented in
Ref.~\cite{BMZ}.  Given $K$ independent and identically
distributed random variables $E_i$, pulled from a distribution of
the form
\begin{equation}
\tilde{P}(E)=\frac{A}{|E|^\alpha} \exp(-B|E|^\delta),
\label{eigendist}
\end{equation}
the probability that the lowest of the $K$ energies is $E$ (for $E
\rightarrow -\infty $ and $K \rightarrow \infty$) is given by
\begin{equation}
P(E) \propto \exp{[B\delta |E_c|^{\delta-1} E]} \ ,
\label{extreme}
\end{equation}
where
\begin{equation}
E_c=-\left( \frac{\log(K)}{B} \right)^{1/\delta}. \label{Ec}
\end{equation}
The value of $E_c$, the lowest energy expected to be attained in K
trials, is easily obtained from
\begin{eqnarray}
\int_{-\infty}^{E_c} dE \tilde{P}(E) \simeq 1/K. \label{Ecform}
\end{eqnarray}

The reason why we chose a distribution of the form given in
Eq.~(\ref{eigendist}) is that in $d=1$ the probability
$\tilde{P}(E)$ is known to have that form exactly for the case of
delta correlated random potentials (see \cite{halperin}).  For
$d>1$ Lifshits \cite {lifshits} argued that the form given by
Eq.~(\ref{eigendist}) is also valid. Our goal now is to see if the
distribution Eq.~(\ref{eigendist2}) derived using the 1-step RSB
solution is indeed consistent with the distribution
Eq.~(\ref{extreme}) predicted using extreme value statistics.

Comparing Eq.~(\ref{eigendist2}) and Eq.~(\ref{extreme}) we find
that for consistency the break point should satisfy
\begin{equation}
x_c=\frac{\delta}{\beta L}
B^{1/\delta}(\log(K))^{(\delta-1)/\delta}.
\end{equation}
Notice that the $1/L$ behavior of $x_c$ is exactly the same as was
found analytically for large $L$ in Ref. \cite{gold} and
numerically for any $L>L_c$ in the present work. We can go further
by using the fact that the number of energy levels $K$, within a
fixed energy interval is directly proportional to the system size,
which in our formulation is effectively determined by $\mu$.
Assuming $\log(K) \propto |\log(\mu)|$ and comparing to the
approximate solution for $x_c$ in Eq.~(\ref{xcf}) we find that
$\delta=(4-d)/2$ and $B \propto 1/g$. Now $\tilde{P}(E)$ is just
proportional to the density of states $\rho(E)$, and it is known
exactly in one dimension. Indeed, when $d=1$, $\delta=3/2$ and $B
\propto 1/g$.
For $2 \leq d < 4$, $\delta$ agrees with the result
derived by Lifshits \cite{lifshits}. Hence, the exponent $\delta$
and the disorder dependence of $B$ is correctly predicted by the
1-step RSB solution.

The above results show that the 1-step RSB solution correctly
predicts some important features of the eigenvalue distribution.
More importantly, we have shown that the 1-step RSB solution can
be interpreted in terms of the eigenstates of the Schr\"odinger
equation with a random potential. However, there are differences
and these reveal the limitations of the 1-step RSB solution. For
example, all the localized states are approximated by the same
Gaussian profile when in fact the localization lengths should
increase with energy.

\section{Localization of polymers in a medium with fixed random obstacles}
In this section we discuss the static properties of a Gaussian polymer
chain without excluded volume interactions that is confined to a
medium populated with quenched random obstacles.
It is important to distinguish between the following two important cases
that have been discussed in the literature:

1. A Gaussian random potential with short range correlations.

2. Random obstacles which prevent the chain from visiting certain sites.

Numerical simulations performed in three dimensions were
restricted, to our knowledge, only to the case of random obstacles
\cite{BM,leung,dayantis}. On the other hand, extensive analytical
work using the replica variational approach and Flory type free
energy arguments, has been done for the case of a  Gaussian random
potential \cite{cates,edwards,shifgold,gold,nattermann}. The case
of a bounded (saturated) random potential was also addressed in
Ref. \cite{cates}. It was not clear to us to what extent these
theoretical investigations could be applied to the case of
infinitely strong random obstacles placed randomly in the medium,
as simulated numerically. This motivated us to investigate this
problem in detail \cite{gs_ro}. The results indicated that only in
a special case (a small value of the embedding volume) the two
problems mentioned above are similar, but otherwise they are quite
different.

We will assume that the obstacles are
infinitely strong--they totally exclude the chain from visiting a given site
occupied by an obstacle. Each obstacle is taken to be a block of volume
$a^{d}$, where $d$ is the number of spatial dimensions and where $a$ is the
linear dimension of the block. We take for simplicity the polymer bond length
$b$ to be approximately equal to $a$. Thus, $a$ will be the small length scale
in the problem, and we will measure all distances in units of $a$.
The obstacles are placed on the sites of a cubic
lattice with lattice spacing $a$. We denote by $x$ the probability that any
given lattice site is occupied by an obstacle (block). Our main results will
concern the case of small $x$, in particular $x<x_{c}$, where $x_{c}$ refers
to the percolation threshold ($x_{c}=0.3116$ for a cubic lattice in $d=3$),
but we will also comment on the case of a larger concentration of obstacles.
We denote by ${\cal V}$ the total volume of the system.

Assume that the chain is occupying a
spherical region (lacuna) of volume $\Omega\sim R^d$. In this region
the actual volume fraction of obstacles will be denoted by $\hat{x}$. It is
crucial to realize that although the average number of obstacles per
site is fixed by $x$, the actual number of obstacle in a small region
of volume $\sim R^d$
 is a fluctuating quantity which occurs with probability $ b(R^d \hat{x};
R^d,x)$, where
\begin{eqnarray}
 b(k;n,p) =\binom{n}{k}p^{k}(1-p)^{n-k},\label{binom}
\end{eqnarray}
  denotes the binomial probability distribution.

In the limit where the system has infinite volume $\cal{V}$ the free
energy for the chain is given by
\begin{eqnarray}
  F(R,\hat{x})=-L \ln(z)+ \frac{L}{R^2}+L\hat{x}-\ln[b(R^d \hat{x};R^d,x)],
\end{eqnarray}
All these terms originate from entropy $F=-TS$ where for simplicity we
take $T=1$ since the temperature does not play a significant role here
with respect to the results. The first term originates from the
entropy of a free chain in $d$-dimensions where $z$ is the
coordination number which for a cubic lattice is equal to $2d$. The
second term originates from the entropy of confinement in a cavity of
radius $R$. The third term is the entropy loss due to obstacles. This
linear dependence is justified in Ref.~\cite{gs_ro} and in the
Appendix therein.
The forth term represents an entropy given by the logarithm of the
probability to have a region of size $\Omega$ with $\Omega \hat{x}$ obstacles.
This free energy, valid for ${\cal
  V}\rightarrow\infty$, is called the annealed free
energy since when the polymer can sample the entire space it is the
same as the random potential adjusting itself to the polymer configuration.
The free energy has to be minimized (the entropy maximized) with
respect to $R$ and $\hat{x}$. The most favorable value of $\hat{x}$ is
0. Since $b(0;R^d,x)=(1-x)^{R^d}$, we find
\begin{eqnarray}
F(R)=-L\ln(z)+\frac{L}{R^{2}}-R^{d}\ln(1-x).
\label{free0}
\end{eqnarray}
This free energy has now to be minimized with respect to $R$ to yield
\begin{eqnarray}
R_{m,annealed}\sim\left(  \frac{L}{|\ln(1-x)|}\right)  ^{1/(d+2)}%
.\label{annealedR}%
\end{eqnarray}
Thus the size of the chain grows with $L$, but with an exponent
smaller than $1/2$, the free chain exponent.

So far we discussed the case of an infinite volume ${\cal V}$. In a finite
volume we find that the so called quenched and annealed case differ, at least
when the volume is not too big. We actually find that there are three regions
as a function of the size of the system volume ${\cal V}$. First, if
${\cal V<V}_{1}\simeq\exp(x^{-(d-2)/2}/(1-x))$, it is unlikely for a chain
of volume $\Omega \sim R^{d}$ to find a region which is totally free of
obstacles. Thus $\hat{x}$ does not vanish in this regime. To proceed further we must
use an approximation to the binomial distribution $b(\Omega
\hat{x};\Omega ,x)$.

If $\Omega$ is not too small
we can approximate the binomial distribution by a normal distribution
\cite{feller}
\begin{eqnarray}
b(\Omega \hat{x};\Omega,x)\approx(2\pi \Omega x(1-x))^{-1/2}\exp\left(
  -\frac{\Omega (\hat{x} -x)^{2}}{2x(1-x)}\right)  .
\end{eqnarray}
This approximation is good provided $\Omega x \gg 1$ and $\Omega
(1-x)\gg 1$. We verified
that these conditions are indeed met in our case when $x$ is small.

In a finite volume ${\cal V}$, the lowest expected value of $\hat{x}$, to be
denoted by $\hat{x}_{m}$, can be found from the tail of the
distribution
\begin{eqnarray}
\int_{0}^{\hat{x}_{m}}d\hat{x}\exp\left(  -\frac{\Omega(\hat{x}-x)^{2}}
{2xy}\right)  \simeq\frac{\Omega}{{\cal V}},
\label{xm}
\end{eqnarray}
which gives
\begin{eqnarray}
\hat{x}_{m}\simeq x-\sqrt{\frac{xy\ln{\cal V}}{R^{d}}},
\end{eqnarray}
where we put $y\equiv 1-x$.
The free energy becomes
\begin{eqnarray}
F_{I}(R)=-L\ln(z)+\frac{L}{R^{2}}+Lx-L\sqrt{\frac{xy\ln{\cal V}}{R^{d}}.}
\end{eqnarray}
The last term in the annealed free-energy is missing since it is negligible for
large $L$ when $R$ is independent of $L$. Minimizing $F(R)$ with
respect to $R$ we find
\begin{eqnarray}
R_{mI}\sim\left(  xy\ln{\cal V}\right)  ^{-1/(4-d)}\label{Rm1}
\end{eqnarray}
and
\begin{eqnarray}
\hat{x}_{mI}=x-\left(  xy\ln{\cal V}\right)  ^{2/(4-d)}.
\label{xmI}
\end{eqnarray}
The result for the radius of gyration of the chain, as represented by $R_{mI}$
is the same result as for the case of the Gaussian distributed random
potential, but with the strength $g$ replaced by $x(1-x)$. The polymer in this
case is localized and its size is independent of $L$ for large $L$.

As ${\cal V}$ grows $R_{m}$ decreases until eventually $\hat{x}_{m}$ vanishes.
This happens when ${\cal V=V}_{1}\simeq\exp(x^{-(d-2)/2}y^{-1})$. For
${\cal V>V}_{1}$, $R_{m}$ is no longer given by $R_{mI}$, but rather
by the solution of $\hat{x}_{mII}=0$. It is the largest region free of
obstacles expected to be found in a volume ${\cal V}$. Rather than
using the normal approximation we can estimate $R_m$ directly from the
relation
\begin{eqnarray}
(1-x)^\Omega \simeq \Omega /{\cal V},
\end{eqnarray}
with $\Omega \sim R_m^d$. Solving for $R_m$ we obtain
\begin{eqnarray}
R_{mII}\sim\left(  \frac{\ln{\cal V}}{|\ln(1-x)|}\right)^{1/d}.
\label{Rm2}
\end{eqnarray}
The polymer is still localized but the dependence on $x$ and on $\ln
{\cal V}$ has changed.
In this region which we call region II the free energy is given by
\begin{eqnarray}
F_{II}=-L\ln(z)+\alpha L\left(  \frac{\ln{\cal V}}{|\ln(1-x)|}\right)
^{-2/d},\label{F2}
\end{eqnarray}
where some undetermined constant $\alpha$ is introduced for later convenience.

As ${\cal V}$ grows in region II, $R_{mII}$ continues to grow until it
reaches the annealed value given above. This happens when
\begin{eqnarray}
{\cal V=V}_{2}\sim\exp\left(  x^{2/(d+2)}L^{d/(d+2)}\right)
\end{eqnarray}
to leading order in $x$, which is enormous for large $L$.
For ${\cal V>V}%
_{2}$ we have the third region in which $R_{mIII}=R_{m,annealed}$ and it grows
like $L^{1/(d+2)}$.

We can thus summarize the behavior of the end-to-end distance as the
function of the system's volume as follows:

Region I   \ \ \ \ ${\cal V}<{\cal V}_1\simeq\exp(x^{-(d-2)/2})$

\begin{eqnarray}
R_{mI}\sim\left(  x\ln{\cal V}\right)  ^{-1/(4-d)}\label{Rm_1}%
\end{eqnarray}
Region II \ \ \ \  ${\cal V}_1<{\cal V}<{\cal V}_2\sim \sim\exp\left(
  x^{2/(d+2)}L^{d/(d+2)}\right)$
\begin{eqnarray}
R_{mII}\sim\left(  \frac{\ln{\cal V}}{|\ln(1-x)|}\right)^{1/d}.
\label{Rm_2}
\end{eqnarray}
region III \ \ \ \  ${\cal V}_2<{\cal V}$

\begin{eqnarray}
R_{m}\sim\left(  \frac{L}{|\ln(1-x)|}\right)  ^{1/(d+2)}%
.\label{annealed_R}%
\end{eqnarray}

The behavior in region II can be deduced from known results of the
density of states for a quantum particle in the presence of obstacles
(repulsive impurities). In that case \cite{lifshits} the
density of states is given by (when the obstacles are placed on a
lattice)
\begin{eqnarray}
\rho(E)\sim\exp(-c|\ln(1-x)|E^{-d/2}),\ E>0
\end{eqnarray}
with $c$ being some dimension dependent constant and $x$ is the density of
impurities. Note that $\rho(E)$ vanishes for $E<0$. We can estimate the lowest
energy in a finite volume ${\cal V}$ from the integral
\begin{eqnarray}
\int_{0}^{E_{c}}dE\rho(E)\simeq1/{\cal V},
\end{eqnarray}
and find
\begin{eqnarray}
E_{c}\sim\left(  \frac{\ln{\cal V}}{|\ln(1-x)|}\right)^{-2/d},
\end{eqnarray}
and thus the localization length is given by
\begin{eqnarray}
\ell_{c}\sim|E_{c}|^{-1/2}\sim\left(  \frac{\ln{\cal
      V}}{|\ln(1-x)|}\right)  ^{1/d}.
\end{eqnarray}

We now make some remarks on the validity of the spherical
droplet approximation. The shape of a long polymer chain is determined by the
regions of the random medium that have a lower than average number of
obstacles. For ${\cal V}>{\cal V}_{1}$ these regions are essentially
free of obstacles. The probability of finding such empty regions depends only
on its volume and not its shape. However given regions of varying shapes and
equal volumes, it will be entropically more favorable for a long polymer chain
to reside in a region whose shape is closest to a sphere. This is because the
confinement entropy is maximized for a sphere over other shapes of the same
volume. The argument is equivalent to that proposed by Lifshits
\cite{lifshits} in the context of electron localization and is shown rigorously
by Luttinger \cite{luttinger}. For ${\cal V}<{\cal V}_{1}$ the relevant
regions contain a small number of obstacles but we believe that the same
argument should roughly hold and deviations from a spherical shape
will be small or irrelevant.

We have compared our analytical results with numerical simulations
performed by Dayantis \textit{et al. }\cite{dayantis}, and also comment on the
relation to earlier simulations done by Baumgartner and Muthukumar \cite{BM}.
Dayantis \textit{et al. } carried out simulations of free chains
(random-flight walks) confined to cubes of various linear dimensions $6-20$, in
units of the lattice constant. These chains can intersect freely and lie on a
cubic lattice. They introduced random obstacles with concentrations $x=0, 0.1,
0.2$
and $0.3$. The length of the chains vary between $18-98$ steps. They also
simulated self-avoiding chains that we will not discuss here. They measured
the quenched entropy, the end-to-end distance, and also the radius of gyration
which is a closely related quantity. Unfortunately, these authors did not have
a theoretical framework to analyze their data, and thus could not make it
collapse in any meaningful way. We show below how it is possible to fit the
data nicely to our analytical results.

\begin{figure}[ptb]
\centering
\includegraphics[width=3.5in,height=3.5in]{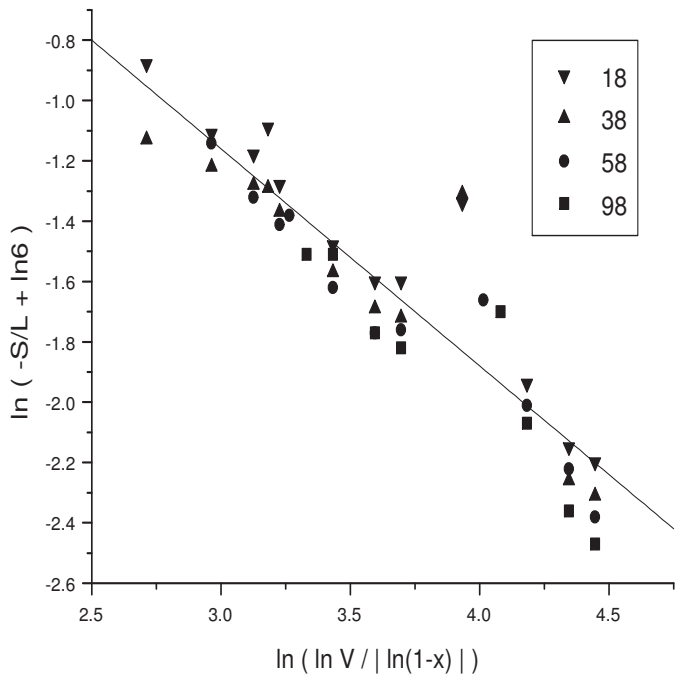}
\caption{A plot
of $\ln(-S/L+\ln 6)$ vs. $\ln(\ln{\cal V}/|\ln(1-x)|)$. The labels are
marked according to the chain length.}
\label{fig:rofe}
\end{figure}

Even for $x=0.1$, the value that we get for ${\cal V}_{1}$ is about $33$
which is an order of magnitude smaller than the the smallest volume used in
their simulation, which is $216$ for a cube of side $6$. Hence we expect to be
in region II. To check the agreement with our analytical results we show in
Figure \ref{fig:rofe} a plot of $\ln(-S/L+\ln6)$ vs. $\ln(\ln{\cal V}/|\ln(1-x)|)$ where $S$ is
the entropy measured in the simulations and ${\cal V}=B^{3}$ for a box of
side $B$. Recall that $F=-S$ and Eq.~(\ref{F2}) predicts a straight line with
slope $-2/3$. The best fit is obtained for a slope of $-0.72\pm0.05$, which is
in excellent agreement with our analytical results in region II.

In order to analyze the simulation results for the end-to-end distance and
radius of gyration we have to introduce some additional compensation for the
results obtained in the previous section. First we must realize that
Eq.~(\ref{Rm2}) is valid only when the number of steps (monomers) is very
large. In the simulations they used chains of varying lengths whose size did
not yet reach asymptotia. Hence, we introduce a correction factor
\begin{equation}
R_{m}(L)=R_{m}(1-\exp(-L/R_{m}^{2}))^{1/2}\equiv R_{m}f_{1}(\sqrt{L}/R_{m})
\ ,
\end{equation}
which interpolates between the size of a free chain as $L\rightarrow0$ and the
value of $R_{m}$ from Eq.~(\ref{Rm2}) as $L\rightarrow\infty$.

\begin{figure}[ptb]
\centering
\includegraphics[width=3.5in,height=3.5in]{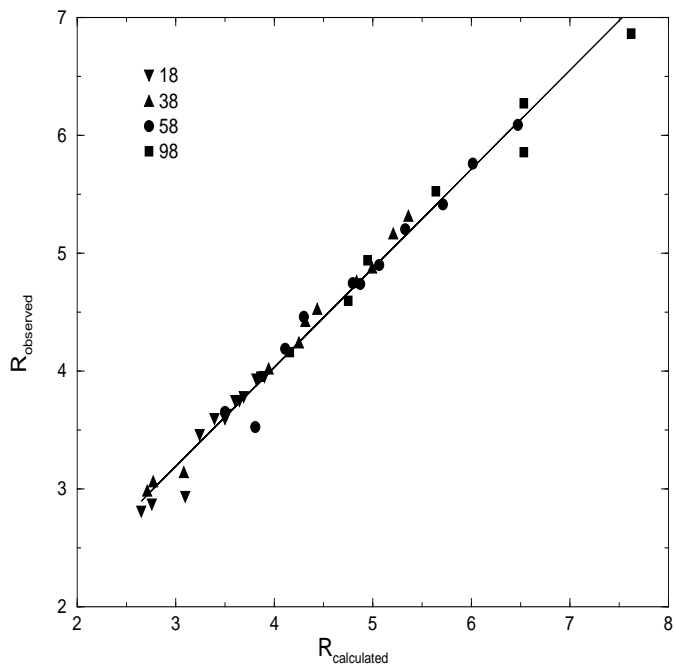}
\caption{A plot of
the observed vs. calculated end-to-end distance}%
\label{fig:roete}%
\end{figure}

The second correction we have to implement arises when the expected value of
the chain is not much smaller than the size of the confining box. Even for a
free chain confined to a box of side $B$ with no obstacles present, the
end-to-end distance is not simply $R=L^{1/2}$ for $L^{1/2}<B$ and $R=B$ for
larger $L$. We have to take into account the fact that the length of the chain
has a Gaussian distribution about its expected value, and the tail of the
Gaussian is cut off by the presence of the box (this is for the absorbing
boundary conditions that is used in the simulations). Thus, for the case of no
obstacles ($x=0$), The measured end-to-end distance should approximately be
\begin{equation}
R_{c}^{2}=\int_{-B}^{B}dR\ R^{2}\exp(-\frac{R^{2}}{2L})/\int_{-B}^{B}%
dR\ \exp(-\frac{R^{2}}{2L}),
\end{equation}
which gives $R_{c}=\sqrt{L}f_{2}(B/\sqrt{L})$ with
\begin{equation}
f_{2}(x)=\left(  1-\sqrt{\frac{2}{\pi}}\frac{x}{\operatorname{erf}(x/\sqrt
{2})}\exp(-x^{2}/2)\right)  ^{1/2}.
\end{equation}
This indeed gives good agreement with the measured values in the no obstacle
case. For the obstacle case we thus have to introduce these two corrections in
subsequent order:
\begin{equation}
R_{m,corrected}=R_{m}f_{1}(\sqrt{L}/R_{m})f_{2}\left(  \frac{B}{R_{m}%
f_{1}(\sqrt{L}/R_{m})}\right)  ,\label{Rcalculated}%
\end{equation}
where $R_{m}=R_{mII}$ as given by Eq.~(\ref{Rm2}). (A constant of
proportionality of $1.8$ has been introduced on the rhs of
Eq.~(\ref{Rm2}) to obtain a good fit).
In Figure \ref{fig:roete} we show
a comparison of the simulation results for the end-to-end distance
with the calculated results as given by Eq.~(\ref{Rm2}) and
Eq.~(\ref{Rcalculated}). The agreement seems remarkable, since all the
data collapses to a straight line with a slope close to $1$.

Dayantis \textit{et al.} emphasize that they did not consider concentrations
of obstacles above the percolation threshold, which is at $x_{c}=0.3116$ for a
simple cubic lattice. The reason is that above the percolation threshold the
medium of random obstacles begins to form disconnected islands free of
obstacles. Thus, in their simulation the polymer chain will only sample a
limited fraction of the volume available. What happens is that effectively the
volume available for the chain is not the total volume of the cube but rather
the volume of the disconnected region it occupies. For most realizations of
the random medium this effective volume will be smaller than the value
${\cal V}_{1}$, which is the limit of region I of the last section. In that
case one expects the end-to-end distance to scale like $x^{-1}$ as given in
Eq.~(\ref{Rm1}) instead of like $x^{-1/3}$ as given by Eq.~(\ref{Rm2}).
Baumgartner and Muthukumar's simulation was for both below the percolation
threshold and also above it ($x=0.4$ and $0.5$). However, they only estimate
the exponent above the percolation threshold, and find it to be about $-1$.
They do not estimate the exponent for $x$ below the percolation threshold,
which appears from their data to scale with a much smaller exponent. Thus, it
seems likely that the reason these authors report a behavior corresponding to
region I, even though their box is quite large, is because the effective
volume is small for the cases for which they exceed the percolation threshold.

\section{Localization and delocalization of polymers with
  self-avoiding interaction in the presence of disorder}

In the previous sections an ideal (Gaussian) chain has been used,
which corresponds approximately to the experimental situation at
the so called $\Theta$-temperature when the solvent effectively
screens the self-avoiding interaction of the chain. In some early
papers \cite{BM,nattermann,thirumalai} there was an attempt to
include the effects of the self-avoiding interaction of the
polymer. These attempts were far from complete. For example in
Ref. \cite{nattermann} it was assumed that the conformation of the
polymer consists of one spherical blob and it was argued that a
quenched random potential is irrelevant for a very long chain when
a self-avoiding interaction is present. In Ref. \cite{thirumalai}
analytical results were obtained for annealed disorder, and
simulations were performed for strictly self-avoiding walks. Ref.
\cite{BM} presents numerical evidence for a size transition of the
polymer as a function of the relative strength of the disorder and
the self-avoiding interaction. The simulations were carried out
for a random distribution of hard obstacles with a concentration
exceeding the percolation threshold. In a recent paper
\cite{gs_sa} we tried to shed more light on this important
problem. We have made use mainly of Flory-type arguments, and
considered both the case of a Gaussian random potential and the
case of randomly placed obstacles. Note that a polymer with
self-avoiding interactions cannot be mapped into a quantum
particle at a finite temperature in a simple manner, because for a
quantum particle there is no impediment to return at a later time
(or Trotter time) to a position it visited previously.

An important point to keep in mind is the strength of the excluded
volume interaction. If one considers a strictly self-avoiding walk
on a lattice (SAW) corresponding to a non-self-intersecting chain,
then the strength of the Edwards parameter $v$ \cite{doi} is fixed
at $O(1)\times a^{3}$ where $a$ is the step size (or monomer size)
and depends only on the type of lattice. On the other hand one can
consider a Domb-Joyce model \cite{domb} where there is a finite
penalty for self overlapping of polymer segments, and then the
strength of $v$ can be varied substantially and reduced
continuously to zero. The interplay between the strength of the
self-avoiding interaction and the strength of the disorder can
then be investigated to a larger extent. Experimentally the
Edwards parameter is given approximately \cite{degennes} by
$v=a^{3}(1-2\chi)$, where $\chi$ is the Flory interaction
parameter, which depends on the chemical properties of the polymer
and the solvent, and on the temperature (and pressure). It takes
the value 1/2 at the $\Theta$-point. The case $\chi=0$ corresponds
to a solvent that is very similar to the monomer. In general good
solvents have low $\chi$ whereas poor solvents have high $\chi$
resulting in $v$ being negative. In the following we will restrict
ourselves to the case of positive $v$, which leads to the more
interesting and non-trivial results.

We now revisit the meaning of the term localization as applied to polymers
in a random medium. Although some authors connect the compact size of the
chain when $L\rightarrow\infty$ with the notion of localization, this is
actually not so. The compact size should be viewed as a separate feature from
the notion of localization. Recall that for a Gaussian chain in an
uncorrelated Gaussian random potential of variance $g$ the chain has
typical size
\begin{equation}
R_{F}\propto(g\ln{\cal V})^{-1/(4-d)},\label{rfg}%
\end{equation}
and the binding energy per monomer is given approximately by
\begin{equation}
U_{bind}/L\sim-(g\ln{\cal V})^{2/(4-d)}.\label{gsg}%
\end{equation}
To insure localization, the binding energy of a chain
$U_{bind}$ has to exceed the translational entropy $\ln\mathcal{V}$. From
Eq.~(\ref{gsg}) this amounts to the condition
\begin{equation}
\ln\mathcal{V}<L(g\ln\mathcal{V})^{2/(4-d)},
\end{equation}
which holds for any $2\leq d<4$ when $L$ is large enough (for $2<d<4$ and any
fixed $L$, the condition can be satisfied for large enough $\mathcal{V}$) .
This condition assures that the polymer will stay confined at a given location
and will not, under a some small perturbation move to a different location.
Thus repeating an experiment or a simulation with the same fixed realization
of the disorder, but with different initial conditions, will result in finding
the polymer situated at the same region of the sample as in a previous
experiment, provided of course one waits enough time (which can be enormous)
for the system to reach equilibrium. We observe that this condition is
satisfied for large enough $L$ provided the binding energy per monomer is
positive. Another interpretation of the inequality given above in the context
of equilibrium statistical mechanics is that the partition sum is dominated by
the term involving the ground state as opposed to the contribution of the
multitude of positive energy extended states. The contribution of these states
is proportional to the volume of the system and thus the inequality above
results from the condition
\begin{equation}
\exp(-LE_{0})>\mathcal{V}.
\end{equation}
What we will see in the following sections is that in the presence of a
self-avoiding interaction, a localization-delocalization transition occurs
when varying the strength of the the self-avoiding interaction for a fixed
amount of disorder or alternatively upon varying the strength of the disorder
for a fixed value of the self-avoiding interaction.

\subsection{A self avoiding chain in a random potential}

Consider first the case of a random potential with a Gaussian
distribution. For simplicity, the discussion in the rest of this section
be limited to three spatial dimensions ($d=3$).
Recall that in the case when there is no self-avoiding interactions the
optimal size of a chain $R_{m}$ is found by minimizing the free
energy $F$ in Eq. \ref{FQ}.  This yields
\begin{equation}
R_{m}\sim\frac{1}{g\log(\mathcal{V})}\equiv\frac{1}{G(\mathcal{V})},
\end{equation}
where we defined the volume dependent disorder strength by $G(\mathcal{V}%
)=g\ln(\mathcal{V})$. Substituting this result in $F$ we obtain
\begin{equation}
F_{m}=-\frac{L}{3R_{m}^{2}}\approx-G(\mathcal{V})^{2}L.
\end{equation}
We see that $-F_{m}/L$ is the binding energy per monomer, and it
is strictly positive, so the polymer is localized. In what follows
we will assume that $g$ is small enough so that $G\ll1$ for the
given system volume, hence $R_{m}\gg 1$, and the chain is not
totally collapsed unless $\mathcal{V}\rightarrow \infty$.

We now add a self avoiding interaction and assume first that it is
small, i.e. $v<<g$, or at least $v<g$. If the chain is still
localized in the same well,
which we will see momentarily not to hold when $L$ is large, then%
\begin{equation}
F=\frac{L}{R_{m}^{2}}-L\sqrt{\frac{g\log(\mathcal{V})}{R_{m}^{3}}%
}+\frac{vL^{2}}{R_{m}^{3}}\approx-G(\mathcal{V})^{2}L+vG(\mathcal{V})^{3}%
L^{2}. \label{fsa}%
\end{equation}
Here, besides assuming that $v$ is small we assume for the moment
that $L$ is not too big so the last term in the free energy,
resulting from the self-avoiding interaction, is small enough so
one does not have to take into account the change in $R_m$ due to
the presence of $v$. If we plot $F$ vs. $L$, we see that it is
lowest when
\begin{equation}
L=L_{m}=\frac{1}{2vG(\mathcal{V})}.
\end{equation}
Thus if $vg\ll1$, we have $L_{m}\gg1$, and
\begin{equation}
F_{m}=-\frac{G(\mathcal{V})}{4v}.
\end{equation}
For $L=2L_{m}$ the free energy vanishes and for larger $L$ it eventually
increases fast like $L^{2}$. We can now verify that if $L$ does not exceed
$L_{m}$ then the approximation used above, assuming that $R_{m}$ does not
change appreciably from its ideal chain value, is justified. If we
differentiate the above $F$ in Eq.\ (\ref{fsa}) with respect to $R_{m}$ we
find%
\begin{equation}
R_{m}\sim\frac{1}{G(\mathcal{V})}\left(  1+\frac{vL}{R_{m}}\right)
^{2}\approx\frac{1}{G(\mathcal{V})}\left(  1+vG(\mathcal{V})L\right)  ^{2},
\end{equation}
again omitting constants of order unity. Thus the correction $vGL$ evaluated
at $L=L_{m}$ is of order unity, and we can still use the value $R_{m}\sim1/G$.

\begin{figure}
[ptb]
\begin{center}
\includegraphics[
height=3in,
width=5in
]%
{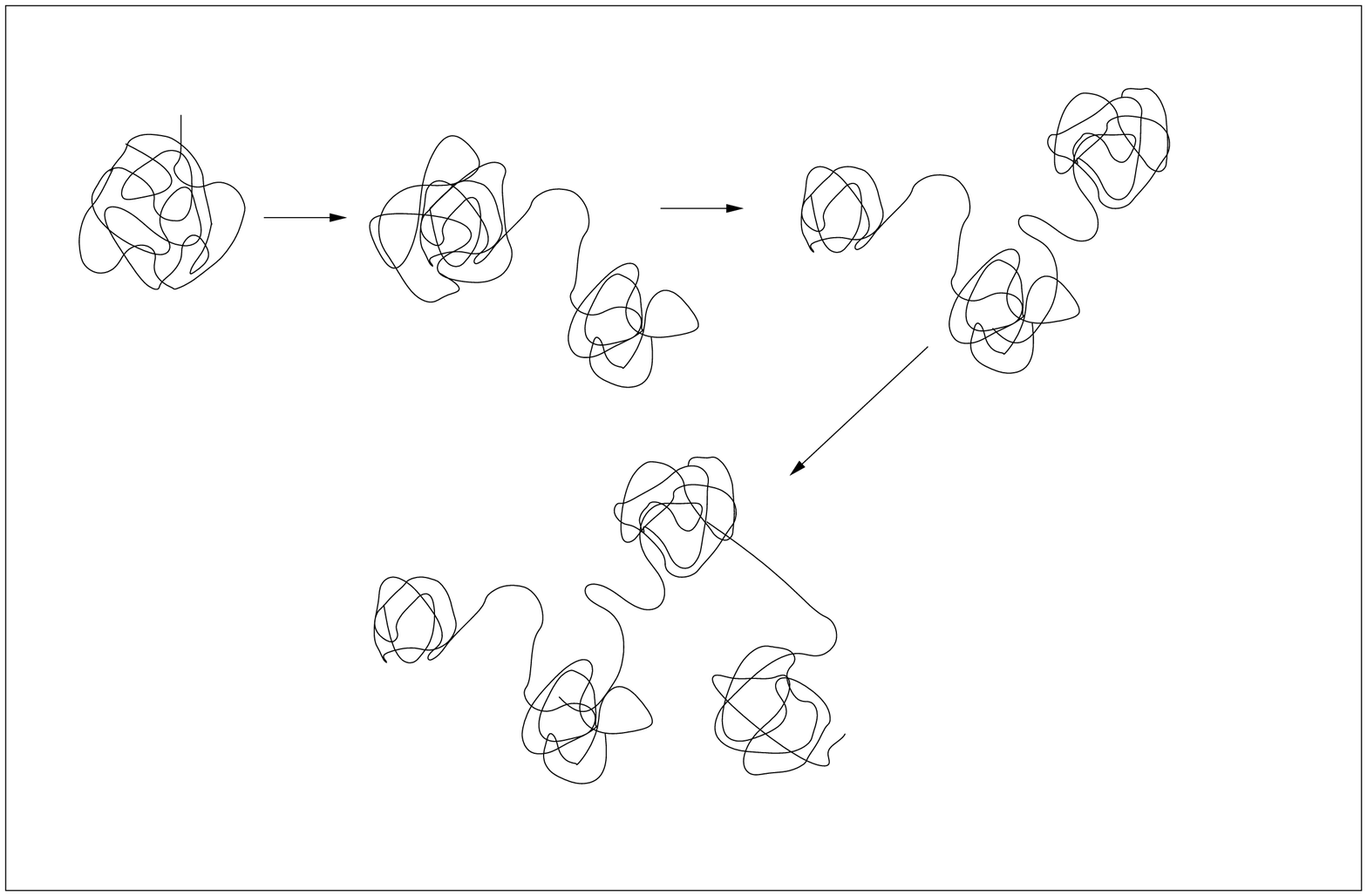}%
\caption{Conformation of a chain consisting of many ``blobs'' with connecting
segments as $L$ increases. Initially the first blob ``overflows'' and there is
``hopping'' to another deep minimum. There is a competition between the length of
the connecting segment and the probability to find a deep minimum nearby.}
\label{fig:blobs}%
\end{center}
\end{figure}

For larger $L$ the approximation seems to break down, but fortunately what
happens is that since the free energy increases when $L$ exceeds $L_{m}$, it
is energetically favorable for part of the chain to jump into a distant well.
Even though there is a cost for the polymer segment between the wells one
still gains in the overall free energy from the binding energy in the wells.
Thus the picture that emerges is that as $L$ increases, the chain divides
itself into separate blobs with connecting segments. In each blob the number
of monomer does not exceed $L_{m}$, which is the optimal value for that well.
The idea is depicted in Figure \ref{fig:blobs}.

To be more specific we will now construct a model for the free energy of the
chain. The first blob will be located in the deepest minimum in the total
volume, whose depth is roughly given by $-G(\mathcal{V})^{2}$ per monomer,
where $\mathcal{V}$ is the total volume of the system. Subsequent blobs will
reside in the most favorable well within a range $Y$, which has to be taken
self-consistently as the length of the jump. Thus within a range $Y$ the chain
is likely to find a potential minimum of depth $-G(Y^{3})^{2}$. The farther
you jump, the deeper well you are likely to find. Thus assuming for simplicity
that the jumps are roughly of equal size, and there are $K$ blobs in addition
to the initial blob, the free energy of the chain will be given roughly by%
\begin{equation}
F(w,m,Y;L)\approx-\frac{G(\mathcal{V})}{4v}+K\left(  \frac{Y^{2}}{m}%
+\frac{m}{Y^{2}}+\frac{vm^{2}}{Y^{3}}-wG(Y^{3})^{2}+vw^{2}G(Y^{3})^{3}\right)
,
\end{equation}
with%
\begin{equation}
K=\frac{L-L_{0}}{w+m},\;L_{0}=\frac{1}{2vG(\mathcal{V})},\;G(Y^{3})=3g\ln(Y).
\end{equation}
We defined $w$ to be the number of monomers in each blob, and $m$ to be the
number of monomers in each connecting segment. The term $Y^{2}/m$ results from
the ``stretching'' entropy of the segment and $m/Y^{2}$ from the confinement
entropy. The term $vm^{2}/Y^{3}$ represents the self-avoiding interaction for
the connecting segments. $L_{0}$ is the number of monomers in the initial blob
whose free energy was taken care of separately. It is evident that when $L$ is
very large we can neglect the free energy of the first blob and also take
$K\approx L/(w+m)$. Thus we find for the free energy per monomer
\begin{equation}
f(w,m,Y)\equiv\frac{F(w,m,Y;L)}{L}\approx\frac{1}{w+m}\left(  \frac{Y^{2}}%
{m}+\frac{m}{Y^{2}}+\frac{vm^{2}}{Y^{3}}-wG(Y^{3})^{2}+vw^{2}G(Y^{3}%
)^{3}\right)  .\label{fepm}%
\end{equation}
This function has to be minimized with respect to $w$, $m$ and $Y$ to find the
parameters giving rise to its lowest value. For the connecting pieces of the
chain we did not include a contribution from the random potential since it is
expected to average out to zero for these parts. To gain some feeling into the
behavior of this function and the values of the parameters which minimize it,
we display in Table I the value of the parameters and free energy per monomer
for $g=0.05$ and various values of $v$, as obtained from a minimization
procedure. The delocalization transition is the point where $f$ changes sign
from negative to positive, as discussed earlier. Actually, to be more precise,
the delocalization transition occurs when $f=-(\ln\mathcal{V})/L$ for finite
$L$, when the translational entropy starts to exceed the binding energy. In
the limit of large $L$ we can say that the transition is at $f=0$. We observe
that the delocalization transition occurs at $v=0.0478$ which is close to the
value of $\ g$. We also observe that $m\ll w$ for $v\ll g$ and $m\gg w$ near
the transition. Also for small $v$, $m\sim Y$, whereas near the transition
$m\sim Y^{2}$. If we compare the value of $w$ from Table I with the value of
$L_{m}=1/(2vG(Y^{3}))$ we find that $w$ is smaller than $L_{m}$ in the entire
range. The ratio $w/L_{m}$ varies from $\sim0.4$ to 1 as $v$ changes from
$10^{-5}$ to 0.048. Thus the assumption we have made previously concerning
this ratio is justified \textit{a posteriori}.

\begin{table}[ptb]
\begin{tabular}
[c]{|l|l|l|l|l|}\hline
v & Y & m & w & f\\\hline
0.00001 & 2206 & 2413 & 16178 & -0.835249\\\hline
0.0001 & 346 & 534 & 2580 & -0.421023\\\hline
0.001 & 60 & 148 & 461 & -0.164673\\\hline
0.01 & 13.7 & 66 & 97 & -0.0370883\\\hline
0.02 & 11.1 & 69.2 & 60.8 & -0.0156444\\\hline
0.03 & 12.8 & 105.5 & 41.9 & -0.0056033\\\hline
0.04 & 22 & 302.6 & 26.8 & -0.0009992\\\hline
0.045 & 35.6 & 720.1 & 20.7 & -0.0001776\\\hline
0.0478 & 69.4 & 2342.4 & 16.5 & 0\\\hline
0.048 & 77 & 2800 & 16 & 0.00000725\\\hline
\end{tabular}
\caption{Parameters and free energy for the case g=0.05}
\label{table1}%
\end{table}

Luckily it was possible to solve the minimization equations analytically
almost entirely in both the limits $v\ll g$, and near the transition when
$f\approx0$. Details of the solutions are given in the Appendix of
Ref. \cite{gs_sa}. Here we only
display the results:

A. The case $v\ll g$.

The parameters are given by%
\begin{align}
Y  &  =\frac{2}{v}(\ln Y-1)^{-1/2}(\ln Y+3)^{-3/2},\label{yvs}\\
m  &  =\frac{2}{3vg\ln Y}(\ln Y-1)^{-1}(\ln Y+3)^{-1},\label{mvs}\\
w  &  =\frac{2}{3vg\ln Y}(\ln Y+3)^{-1}\label{wvs}\\
f  &  =-9g^{2}(\ln Y)^{2}(\ln Y-1)(\ln Y+3)^{-1}. \label{fvs}%
\end{align}
The first equation can be easily solved numerically for $Y$ for a given value
of $v$ and the result substituted in the other equations. Very good agreement
is achieved with Table I for small values of $v$.

B. Solution near the delocalization transition.

Let us define the parameter%
\begin{equation}
\kappa=\frac{4v}{3g}. \label{kappa}%
\end{equation}
In terms of this parameter we have%
\begin{align}
Y  &  =\frac{1}{v\kappa}\left(  \frac{1+\sqrt{1+\kappa^{2}}}{\kappa}\right)
^{2},\label{ytr}\\
m  &  =\frac{1}{v^{2}\kappa^{2}}\left(  \frac{1+\sqrt{1+\kappa^{2}}}{\kappa
}\right)  ^{3},\label{mtr}\\
w  &  =\frac{\kappa}{8v^{2}\ln Y}=\frac{1}{2vG(Y^{3})}. \label{wtr}%
\end{align}
The transition point is obtained by solving the equation%
\begin{equation}
\frac{1+\sqrt{1+\kappa^{2}}}{\kappa}+\frac{\kappa}{1+\sqrt{1+\kappa^{2}}%
}+\frac{1}{\kappa}\left(  1-2\ln\left(  \frac{2(1+\sqrt{1+\kappa^{2}})}%
{\sqrt{3g}\kappa^{2}}\right)  \right)  =0. \label{feqz}%
\end{equation}
Once the solution $\kappa_{c}$ is determined for a given $g$, then the
transition point $v_{c}$ is determined from $v_{c}=3g\kappa_{c}/4$. For $g$ in
the range 0.01-0.2 we find that $\kappa_{c}$ is a number of order unity
(varies from 1.7 to 0.96 as $g$ changes in that range). This means that
$v_{c}$ is quite close to $g$. Once $v_{c}$ is known, all the parameters $Y,m$
and $w$ at the transition are determined by the solution above. For $g=0.05$
we get $\kappa_{c}=1.2744$, and thus $v_{c}=0.0478$ in excellent agreement
with the minimization results from Table \ref{table1}.

\begin{figure}
[ptb]
\begin{center}
\includegraphics[
height=3.3036in,
width=4.5316in
]%
{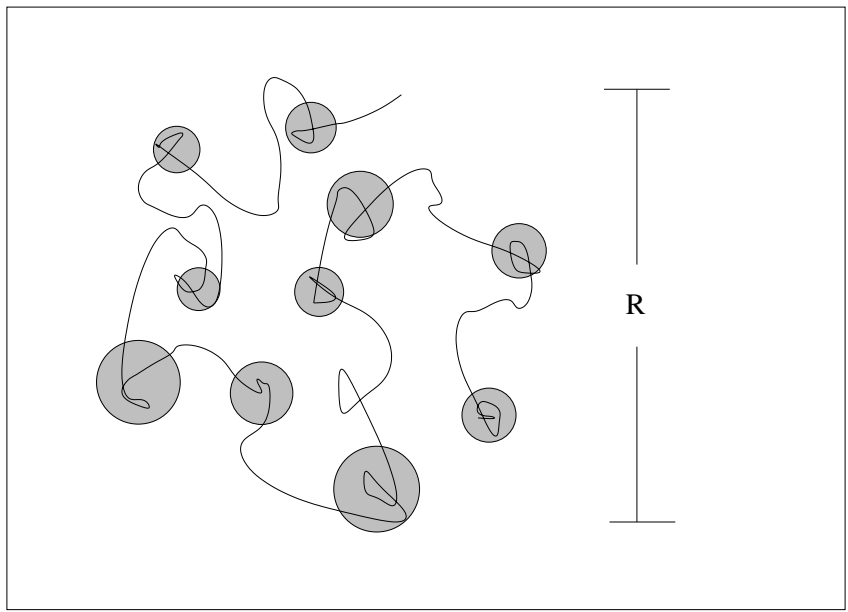}%
\caption{A typical chain conformation when $v\gg g$. The dark regions are
regions of low average potential. Only short segments of the chain are
situated in these regions.}%
\label{fig:largev}%
\end{center}
\end{figure}

For $v>v_{c}$ the chain is delocalized. The above expression for the free
energy may no longer be accurate, but the general picture is clear. There will
be very few monomers in the low regions of the potential, and the chain will
behave very much like an ordinary chain with a self-avoiding interaction in
the absence of a random potential. Any little perturbation can cause the chain
to move to a different location in the medium (see Fig. (\ref{fig:largev})).

For small values of $v$ when the chain is localized, we still expect its size
to grow like that of a self-avoiding walk. Thus we expect roughly
\begin{equation}
R_{g}\sim YK^{0.6},
\end{equation}
since $Y$ is the step size, and $K$ is the number of steps. But since
$K\approx L/(m+w)$ for large $L$, we find%
\begin{equation}
R_{g}\sim\frac{Y}{(m+w)^{0.6}}L^{0.6}.
\end{equation}
Thus the chain behaves as a self avoiding walk with an effective step size per
monomer given by
\begin{equation}
\frac{a_{eff}}{a}=\frac{Y}{(m+w)^{0.6}}.
\end{equation}
From Table \ref{table1} it becomes clear that the effective monomer size changes from a
value of $6$ for $v=0.00001$ to a value of $\sim0.66$ at the transition. The
reason for the large value of the effective monomer size at very small $v$ is
that the chains makes long jumps to take advantage of deep wells of the random
potential, very much like anchored chains in a random potential that make
sub-ballistic jumps \cite{cates}. This is also the reason why the number of
monomers $w$ in each well is somewhat less than $L_{m}$ , since a sufficient
number of monomers need to be used for the connecting segments. For $v\gg v_{c}$
in the delocalized phase we expect the effective monomer size to be about 1,
since the chain behaves almost like an ordinary self-avoiding chain with the
random potential not playing any significant role. Thus the chain is expected
to have its smallest size in the vicinity of the transition.

It is important to notice that the discussion above is in the limit for very
large $L$. If $L\lesssim1/(2vG(\mathcal{V}))$, then in the localized phase
when $v\ll v_{c}$ the polymer will be confined to a single well and will
appear compact, even though it will not remain so for large $L$. This may
explain why in simulations that were done typically with $L\leq320$ \cite{BM}
the delocalization transition appeared as a transition from a compact to a
non-compact state of the polymer.

We should also note that for an annealed random potential there is a
transition from a collapsed state into an ordinary self-avoiding chain as $v$
increases through the point $v=g$ \cite{nattermann}. This is because the
annealed free energy reads%
\begin{equation}
F(R)=\frac{L}{R^{2}}-\frac{gL^{2}}{R^{3}}+\frac{vL^{2}}{R^{3}}+\frac{R^{2}}%
{L}.
\end{equation}
For $v<g$ the fourth term is negligible and the free energy is lowest when
$R\rightarrow0$ (when $L$ is large). For $v>g$ the first term is negligible,
and the radius of gyration grows like
\begin{equation}
R\sim(v-g)^{1/3}L^{3/5}.
\end{equation}

\subsection{A self avoiding chain in a sea of hard obstacles}

We now turn to the case of hard obstacles. This  case of was discussed
at length in the previous section in
the absence of a self-avoiding interaction. We will make use of the results
applicable to  three spatial dimensions. Three different
behaviors were identified as a function of the system's volume. Region I is
defined when the system's volume $\mathcal{V}<\mathcal{V}_{1}\sim\exp
(c_{1}/\sqrt{x})$, where $c_{1}$ is a constant
of order unity. Here $0<x<1$ is
the average concentration of obstacles per site (total number of obstacles
divided by total number of sites). We also assume that $x$ is less than the
percolation threshold ($x_{c}=0.3116$ for a cubic lattice), so sites occupied
by obstacles don't percolate. In Region I we recall from the previous section that in
the absence of a self-avoiding interaction, the free energy per site for a
chain situated in a spherical region of volume $R^{3}$ in three dimensions is
given by
\begin{equation}
F_{I}/L=-\ln(z)+1/R^{2}+\hat{x}%
\end{equation}
where $\hat{x}$ is the actual concentration of obstacles in that region, whose
minimal expected value in a system of total volume $\mathcal{V}$ is
\begin{equation}
\hat{x}_{m}\simeq x-\sqrt{\frac{x\ln\mathcal{V}}{R^{3}}}.
\end{equation}
The binding energy per monomer inside the blob resulting from the lower
concentration of obstacles in this region is given by $x-\hat{x}$, since it is
equal the entropy gain from a lower concentration as compared to the average
(background) concentration $x$. The chain is ``sucked'' towards regions with
low concentration of obstacles since it can maximize its entropy there, and
these regions of space act like the negative potential regions of the Gaussian
random potential. Thus in order for the free energy per monomer to reflect
correctly the binding energy of the chain inside the blob, both the constant
$x$ of the background and the constant term $-\ln(z)$, which is always there
regardless of the chain's position, have to be subtracted. The relevant free
energy per monomer situated in the blob (which is equal to minus the binding
energy) is given by
\begin{equation}
f_{I}=\frac{1}{R^{2}}-\sqrt{\frac{x\ln\mathcal{V}}{R^{3}}}.
\end{equation}
This result coincides with Eq. (\ref{FQ}) upon the substitution
$g\rightarrow x$. Thus all the results of the previous section carry on to
Region I with this simple substitution .

Therefore we are going to discuss the situation when the system's volume is
greater than $\mathcal{V}_{1}$ (Region II). In this case the many blob picture
still holds, where the blobs are now situated in regions free of obstacles
(with $\hat{x}=0$ ) whose size is determined again by the distance $Y$ of the
jump which is also assumed to satisfy $Y^{3}\gtrsim\mathcal{V}_{1}$ ( an
assumption which will be justified \textit{a posteriori}). In this case the
farther the jump, it is more likely for the chain to find a larger space empty
of obstacles, which will reduce further its confinement entropy $w/R^{2}$ and
also the self-avoiding energy ( but there is a cost resulting from the
connecting segments and the constraint of the total length being fixed). The
blob size is given by $R_{mII}=1/G_{o}$ (the largest expected empty region in
a volume $Y^{3}$ ) with \cite{gs_ro}
\begin{equation}
G_{o}(Y^{3})\approx\left(  \frac{x}{3\ln Y}\right)  ^{1/3}.
\end{equation}
Thus the free energy per monomer of a chain consisting of a number of blobs
with connecting parts, each of length $Y$,  is given by%
\begin{equation}
f(w,m,Y)\approx\frac{1}{w+m}\left(  \frac{Y^{2}}{m}+\frac{m}{Y^{2}%
}+\frac{vm^{2}}{Y^{3}}-w(x-G_{o}(Y^{3})^{2})+vw^{2}G_{o}(Y^{3})^{3}\right)  ,
\end{equation}
where again the constant $x$, which is independent of $m$, $w$ and $Y$, has been
subtracted. This assures that the delocalization transition again occurs at
$f=0$ and not at $f=x$. The results of a numerical minimization of this free
energy is displayed in Table \ref{table2} for the case of $x=0.1$.

\begin{table}[ptb]%
\begin{tabular}
[c]{|l|l|l|l|l|}\hline
v & Y & m & w & f\\\hline
0.00001 & 552 & 2206 & 72092 & -0.062\\\hline
0.0001 & 127 & 565 & 9135 & -0.051\\\hline
0.001 & 38 & 206 & 1241 & -0.034\\\hline
0.01 & 23 & 230 & 207 & -0.0077\\\hline
0.02 & 32 & 536 & 137 & -0.0019\\\hline
0.03 & 59 & 1737 & 120 & -0.00035\\\hline
0.04 & 194 & 13915 & 131 & -0.0000099\\\hline
0.041 & 248 & 21327 & 134 & -0.0000023\\\hline
0.042 & 357 & 39310 & 144 & 0.0000024\\\hline
\end{tabular}
\caption{Free energy and parameters for a polymer in a sea of blockers when
$\mathcal{V}>\mathcal{V}_{1}$.}%
\label{table2}
\end{table}

The transition occurs between $v=0.041$ and $0.042$. Again we could find
analytically almost the entire solution both for $v\ll x$ and at the
transition. The solution is given in the Appendix. There we show that the
transition occurs at $v=0.04142$. We observe that as for the case of a
Gaussian random potential the ratio $w/m$ changes from $\gg1$ to $\ll1$ as $v$
approaches the transition from below. The values of $Y$ are seen to be
consistent with the assumption $Y^{3}\gtrsim\mathcal{V}_{1}$ . We also checked
that the free energy from Table II is lower than what one would obtain by
constraining $Y$ to be in Region I, i.e. $Y^{3}<\mathcal{V}_{1}$.

Finally for $\mathcal{V>V}_{2}$, where $\mathcal{V}_{2}\simeq\exp\left(
x^{2/(d+2)}L^{d/(d+2)}\right)  $ (Region III), we expect
the behavior of the
chain to stay the same. This is because for jumps within a volume
$\mathcal{V}_{2}$ the situation reverts to the previously discussed scenario,
since the effective volume of interest which determines the statistics of the
free spaces is of order $Y^{3}$ and we don't expect $Y$ to be that large.

An important point to note is that if one performs a simulation with strict
self-avoiding-walks on a diluted lattice (with $x<1)$, one has $v\sim1$, and
hence one will always be in the delocalized phase and will not see any
localization effects \cite{barat}.

A few words are in order about the case of an annealed potential. This case
has been already investigated in the literature \cite{thirumalai}, and we will
review it briefly. The free energy in the annealed case reads (for $d=3$)
\begin{equation}
F(R)\approx\frac{L}{R^{2}}+xR^{3}+\frac{vL^{2}}{R^{3}}.
\end{equation}
The second term represents the entropy cost of a fluctuation in the density of
obstacles that creates an appropriate spherical region of diameter $\sim R$.
It was assumed that the chain occupies a spherical volume, or at least
deviations from a spherical shape are not large \cite{thirumalai}. In the case
$v=0$ one obtains by minimizing the free energy that $R\sim(L/x)^{1/5}$, a
well known result. For $v>0$ the first term is irrelevant (and so is the
``stretching term'' of the form $R^{2}/L$) and one finds that $R\sim
(v/x)^{1/3}L^{1/3}$.
In $d$ dimension, the size scales like $L^{1/d}$ when
$v>0$, which is larger than the $L^{1/(d+2)}$ dependence in the $v=0$ case.
There is no indication for a phase transition in these arguments, although some
authors \cite{thirumalai} speculate that it breaks down for large $v$ and a
transition to a Flory $L^{3/(d+2)}$ dependence takes place.

\section{Conclusions}
In this chapter we have demonstrated the rich behavior of polymer
chains embedded in a quenched random environment.  As a starting
point, we considered the problem of a Gaussian chain free to move
in a random potential with short-ranged correlations. We derived
the equilibrium conformation of the chain using a replica
variational ansatz, and highlighted the crucial role of the system's
volume. A mapping was established to that of a quantum particle in a
random potential, and the phenomenon of localization was explained
in terms of the dominance of localized tail states of the
Schr\"odinger equation.  We also gave a physical interpretation of
the 1-step replica-symmetry-breaking solution, and elucidated the
connection with the statistics of localized tail states. Our
concusions support the heuristic arrguments of Cates and Ball, but it
starts with the microscopic model.

We then proceeded to discuss the more realistic case of a chain
embedded in a sea of hard obstacles.  Here, we showed that the
chain size exhibits a rich scaling behavior, which depends
critically on the volume of the system.  In particular, we showed
that a medium of hard obstacles can be approximated as a Gaussian
random potential only for small system sizes. For larger sizes a
completely different scaling behavior emerges.

Finally we considered
the case of a polymer with self-avoiding (excluded volume)
interactions. In this case it was found that when disorder is
present, the polymer attains the shape like that of a pearl
necklace, with blobs connected by straight segments.  Using Flory
type free energy arguments we analyzed the statistics of these
conformational shapes, and showed the existence of
localization-delocalization transition as a function of the
strength of the self-avoiding interaction.

The work described in this chapter is concerned with static
(equilibrium) properties of polymers in random media. There is a lot
of theoretical work still to be done related to the dynamics, and especially
nonequilibrium properties of polymers in random media. This is also of
practical importance, for example for the separation of chain of
different length or mass like DNA molecules under the effect of an
applied force when embedded in a random medium like a gel \cite{viovy}.

\section{Acknowledgments}
This work was supported by the US Department of Energy (DOE), Grant
No. DE-FG02-98ER45686.

\end{document}